\documentclass[11pt]{article}
\usepackage{geometry}                
\usepackage{graphicx}
\usepackage{amssymb}
\usepackage{epstopdf}
\usepackage{feynmf}
\usepackage{amsmath}
\usepackage{subfig}
\begin{document}
\begin{titlepage}
\begin{flushright}
LU-TP 11-28 \\
September 2011
\end{flushright}

\begin{center}
\textbf{Bachelor Thesis\\}
\vspace{1cm}
\huge{\textbf{Jet softening and decollimation in heavy-ion collisions\\}}
\vspace{1cm}
\large{Viktor Linders\\}
\vspace{1cm}
Department of Astronomy and Theoretical Physics \\
Lund University \\
S\"olvegatan 14A, SE-223 62 Lund

\vspace{1cm}
\large{Supervised by: Konrad Tywoniuk}

\end{center}

\setlength{\unitlength}{1mm}

\vspace{3cm}

\begin{abstract}
Two suggested models of jet fragmentation in a quark-gluon plasma have been tested, combined and further developed. This has been done by generating hard processes in proton-proton collisions at $\sqrt{s}=2.76$ GeV in PYTHIA 8 with a modified parton shower algorithm. Subsequently, a jet analysis of the final state hadrons have been performed with the FastJet recombination package. The results have been compared to unmodified jets in proton-proton collisions at the same collision energy, and to experimental data obtained by ATLAS. It has been shown that modifications of the parton splitting kernels alone is an insufficient modification to reproduce experimental data on dijet energy asymmetry and azimuthal decorrelation. Additional jet decollimation qualitatively reproduces the main features of the data provided certain variables are chosen aptly. We also study other jet characteristics with the model, such as the jet substructure, that may be observable experimentally in a near future.
\end{abstract}

\end{titlepage}

\section{Introduction}\label{sec:Introduction}
\begin{figure}
\begin{center}
\includegraphics[width=3.5in]{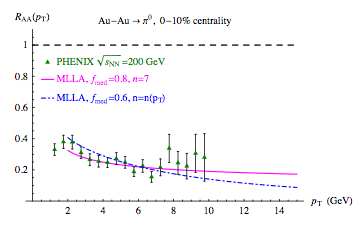}
\caption{Factor 5 supression of the $\pi^0$ meson yield in heavy-ion collisions. Data, taken from Ref. \cite{Kniehl}, is compared to a model for radiative energy loss \cite{Borghini}.}
\label{fig:RAA}
\end{center}
\end{figure}

Quantum chromodynamics (QCD) is the theory of quarks, gluons (jointly referred to as partons) and their interactions. It bears similarities to quantum electrodynamics (QED), not least in the existence of charge. Whereas there is a single QED charge (electric), the  QCD charge comes in three varieties referred to as colors (e.g. red, green and blue) and is carried by quarks. Similarly anti-quarks carry the corresponding anti-color. Gluons, of which there are eight, are the force particles of QCD. In contrast to photons, gluons are not charge neutral but carry a combination of color and anti-color, meaning that gluons are allowed to self-interact. \\
\indent Another remarkable difference between QED and QCD is the coupling, $\alpha_s$, which depends on the momentum scale used. At small momentum scales $\alpha_s$ rapidly becomes large (confinement) whereas at large momenta it tends to zero (asymptotic freedom) \cite{Gross,Politzer}. The QED coupling has an opposite trend, however, can in most cases be considered constant. \\
\indent The concept of asymptotic freedom has led to the realization that at very high temperatures or densities interacting partons become free and transform into a deconfined phase of matter - the quark-gluon plasma (QGP) \cite{Cabibbo,Collins}. The matter of the universe occupied this state a few microseconds after the Big Bang. At present not many details are known about the dynamics of this exotic state of matter. \\
\indent The appropriate conditions for the formation of a QGP can be obtained in a laboratory, namely in high energy collisions of heavy nuclei. This provides the possibility to study its properties in a controlled environment. Attempts in the 1980s and 1990s at SPS (CERN) led to the announcement of indirect evidence for the creation of the new state of matter \cite{CERN}. Formation of a QGP was confirmed at RHIC (2005) \cite{RHIC}, where studies are currently carried out using Au-Au collisions at $\sqrt{s}=200$ GeV per nucleon. The recent completion of the ALICE, ATLAS and CMS detectors in the Large Hadron Collider (CERN) has led to new data gathered from Pb-Pb collisions at $\sqrt{s}=2.76$ GeV per nucleon, i.e. significantly higher than at RHIC. \\
\indent The data indicates that the bulk of the particle production originates from a hot source in thermal equilibrium. One of the most striking features is the so called "jet-quenching" phenomenon. This manifests itself in the suppression of particles with high transverse momenta at RHIC, as seen in Fig. \ref{fig:RAA}, and the LHC. The yield of pions created in head-on heavy-ion collisions is suppressed by a factor $\sim 5$ compared to the expectation, had the nuclear collision simply been a mere superposition of proton-proton collisions. This implies that energetic partons transversing a dense QGP lose significant fractions of their energy, predominantly from radiative processes \cite{ATLAS,CMS}.\\
\indent The purpose of this investigation is to qualitatively reproduce some of the obtained data by constructing a basic model of the parton behavior in a QGP.  Comparing it to similar data for proton-proton collisions, we thus hope to develop a better understanding for the plasma itself and the interactions between it and the particles traversing it. This discussion will mainly focus at the results obtained at the ATLAS experiment at the LHC. \\
\indent We will start with a brief discussion of the running of the strong coupling and how this results in the idea of a quark-gluon plasma. Then follows an outline of the underlying theory and mathematics relevant to factorization of hard processes, culminating in the DGLAP equation. We then proceed to discuss the importance of jets; jet fragmentation; and hadronization in the Lund model, and how to define jets for further implementation. The main investigation then follows with a set of reference results for intrajet distributions, energy distributions and jet energy asymmetry obtained for proton-proton collisions in a vacuum. We then introduce two ideas for modifications that are believed to reproduce data on jets in a QGP obtained at the LHC. The previous distributions are obtained again with the new modifications, followed by a discussion of parameter values relevant to the models. We finish the investigation with a summary of the results and what conclusions can be drawn from them.

\section{Theory}\label{sec:Theory}
\subsection{The running coupling and perturbative QCD}\label{sec:running}
\indent There are two main ways of solving QCD: on the lattice; and using the perturbative approach. The former involves discretizing time and space and considering the values of the quark and gluon fields at all the vertices of the resulting 4-dimensional lattice. This approach is suitable for calculating static quantities in QCD. However, the following discussion will consider high energy systems where lattice QCD is not suitable due to the large number of lattice points required for a complete solution. Instead perturbative QCD will be advised. \\
\indent Perturbative QCD is the idea of an orderly expansion in a small coupling $\alpha_s=\frac{g_s^2}{4\pi} \ll 1$, where $g_s$ is the strong coupling constant. Some observable $f$ can be calculated as

\begin{equation}
f=f_o + f_1\alpha_s + f_2\alpha_s^2 + f_3\alpha_s^3 + \ldots
\label{perturbation}
\end{equation}
where a fixed number of terms only need to be considered as the remaining ones should be negligible, assuming that the $f_i$-terms do not compensate the smallness of $\alpha_s$. \\
\indent In high order QCD calculations it is often convenient to introduce a renormalization scale $\mu$, in order to preserve consistent dimensions for all quantities. The coupling constant $\alpha_s$ depends on the scale at which it is evaluated: 

\begin{equation}
\alpha_s(\mu^2) = \frac{1}{b_0\ln\frac{\mu^2}{\Lambda_{QCD}^2}}, \qquad \alpha_s(Q^2) = \frac{1}{b_0\ln\frac{Q^2}{\Lambda_{QCD}^2}}
\label{coupling}
\end{equation}
In eq. (\ref{coupling}) the coupling is expressed in terms of the renormalization scale and in terms of a more conventional momentum scale $Q$ applied to collider events. Here $\Lambda_{QCD}$ is a constant scale at which the coupling diverges, and for which, if $\mu \gg \Lambda$ (or $\alpha_s(\mu^2) \ll 1$), perturbative QCD is valid, and $b_0$ is (almost) a constant \cite{Gross}\cite{Politzer}. It is suggested from eq. (\ref{coupling}) that at low momentum scales quarks are strongly bound, i.e. \emph{confined} into hadrons, whereas at high momentum scales the coupling becomes weak and the quarks behave as if they were \emph{asymptotically free}. \\
\begin{figure}
\begin{center}
\includegraphics[width=2.5in]{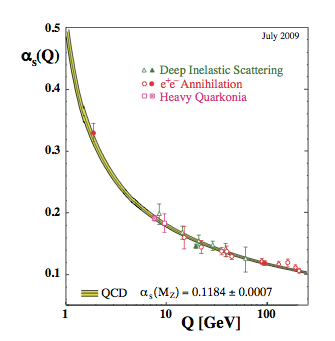}
\caption{The strong coupling $\alpha_s$ runs with the momentum scale $Q$ in an inverse logarithmic manner. Figure provided by Ref. \cite{Bethke}}
\label{running}
\end{center}
\end{figure}
\indent As mentioned previously, asymptotic freedom has led to the idea that as some sufficiently high momentum scale the strong coupling should be weak enough to allow a deconfined state of partons. It is sometimes useful to express the strong coupling as $\alpha_s(k_BT)$. Various calculations have been done in the attempt to find the critical transition temperature where the partons become deconfined \cite{Hagedorn,Karsch}. This is an example of a calculation where QCD on the lattice is a suitable approach, and is also employed by Ref. \cite{Karsch}. It is believed that the transition temperature is of the order $\mathcal{O} (100)$ MeV corresponding to more than a trillion Kelvin. This temperature range would have been reached a few microseconds after the Big Bang, after which the resulting quark-gluon plasma would have cooled down below the critical temperature and hadronization would have begun, forming the commonplace hadronic matter of every day life. \\

\subsection{Factorization properties}\label{sec:crosssections}
\begin{figure}
\begin{center}
\includegraphics[width=4in]{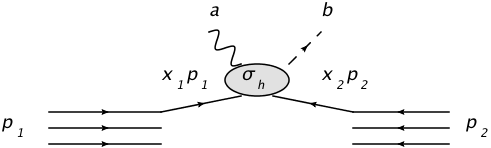}
\caption{Cross section for some arbitrary hard process $q \bar{q}\rightarrow ab $}
\label{fig:crosssection}
\end{center}
\end{figure}

In predicting the probability of any event to occur it is common practice to consider the cross section, $\sigma$, of the event. In the simplest of cases one may consider two incoming particles in a head on collision resulting in two outgoing particles. The cross section of such an event would depend on the momentum fractions of the incoming particles, $x_i$ ($i=1,2$), carried by the colliding quarks, and the total energy transfer in the collision. For processes at high energies, it factorizes into the following form, cf. Fig. \ref{fig:crosssection}, namely

\begin{figure}
\begin{center}
\includegraphics[width=4in]{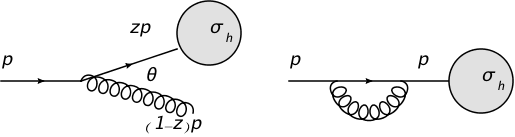}
\caption{(left) Initial state parton splitting. (right) Virtual term.}
\label{fig:cross}
\end{center}
\end{figure}

\begin{equation}
\sigma_h = \int \mathrm{d}x_1f_{q/p}(x_1) \int \mathrm{d}x_2f_{\bar{q}/p}(x_2) \sigma_{(q\bar{q}\rightarrow ab)}(x_1x_2s).
\label{simple}
\end{equation}
Here $s$ is the square of the center of mass energy such that the partonic cross section depends on the energy transfer $(x_1x_2s)$. The subscript $h$ has been added to emphasize that the process is hard. The function $f_{q/p}(x_1)$ is the number density of quarks of type $q$ carrying a fraction $x_1$ of the momentum $p_1$ of one of the incoming baryons (and similarly $f_{\bar{q}/p}(x_2)$ for the other one). These functions are referred to as parton distribution functions (PDFs) and encode the long-distance dynamics of the process. The partonic cross section, $\sigma_{(q\bar{q}\rightarrow ab)}$, on the other hand describes the short-distance process. \\
\indent Quarks being charged with color may well end up radiating a gluon at some stage in the event. It is thus of importance to consider what corrections for $\sigma_h$ would be required in such an occasion. The following line of argument will consider only the case of an initial state emission (Fig. \ref{fig:cross}, left), however it should be noted that equivalent conclusions would be reached if final state emission was regarded instead.

\begin{figure}
\begin{center}
\includegraphics[width=3in]{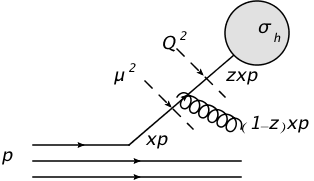}
\caption{The correction to the hard cross section depends on the momentum scale $Q$ and the factorization scale $\mu_F$.}
\label{fig:total}
\end{center}
\end{figure}

The correction to the cross section is certainly going to depend on the momentum fraction of the incoming quark, $z$, carried by the emitted gluon. However, an additional complication enters the picture in that the probability of emission depends on the momentum (resolution) scale, $Q$, of the process.

\begin{equation}
\sigma_{g+h}( p)\simeq \sigma_h( zp)\frac{\alpha_sC_F}{\pi}\frac{\mathrm{d}z}{1-z}\frac{\mathrm{d}k_t^2}{k_t^2}, \qquad C_F=\frac{N_C^2-1}{2N_C}=\frac{4}{3}
\label{emission}
\end{equation}
Here $z$ is the momentum fraction the quark attains from the incoming parton, $N_C=3$ is the number of QCD colors, and $k_t=(1-z)p\sin \theta$ is the transverse momentum of the gluon with respect to the axis of the incoming beam. In the first term after the equating sign in eq. (\ref{emission}) it should be noted that the hard cross section is now dependent on the remaining momentum of the quark, $zp$. The following term describes the strength of the interaction. The penultimate term relates to the momentum fraction, $(1-z)$ carried by the gluon. Note that the cross section diverges for $z\rightarrow1$. This is an important result and suggests that the gluon should carry a comparably small momentum i.e. the gluon is \emph{soft}.\\
\indent The final term of eq. (\ref{emission}) describes the momentum carried by the gluon in a direction perpendicular to the original axis. It is customary to employ transverse variables to avoid biases from boosts in colliders. Notice that the expression diverges for $k_t\rightarrow0$, thus suggesting that the emitted gluon should carry very little transverse momentum i.e. emission should occur at a small angle compared to the beam axis. This is referred to as \emph{collinear} emission. \\
\indent Certainly it is not wished for to have cross sections diverging to infinity. It turns out that if virtual terms are included, such as Fig. \ref{fig:cross} (right), the soft divergence is cancelled out. To resolve the issue of collinear divergence a lower integration limit, $\mu_F$, is introduced. This cutoff, referred to as a factorization scale, creates a limit $k_t < \mu_F$ below which all gluon emission are absorbed by the PDF from eq.(\ref{simple}). As a consequence the PDF becomes dependent on the cutoff (Fig. \ref{fig:total}) and the total correction to the cross section may be written

\begin{equation}
\sigma_{h+g+V} \simeq \frac{\alpha_sC_F}{\pi}\int_{\mu_F^2}^{Q^2}\frac{\mathrm{d}k_t^2}{k_t^2}\int\frac{\mathrm{d}x\mathrm{d}z}{1-z}\left[\sigma_h(zxp)-\sigma_h(xp)\right]q(x,\mu_F^2)
\label{correction}
\end{equation}
where $V$ represents the virtual term. The longitudinal momentum fraction $x$ of the quark originating from the incoming baryon has been included. All gluon emission of $k_t < \mu_F$ are absorbed by the PDF and exclusively those of larger transverse momentum are explicit in the $\mathcal{O}(\alpha_s)$ term in eq. (\ref{correction}). This term is now finite, however possibly large if $Q \gg \mu_F$. 

\subsection{DGLAP evolution}\label{sec:DGLAP}
\begin{figure}
\begin{center}
\includegraphics[width=5in]{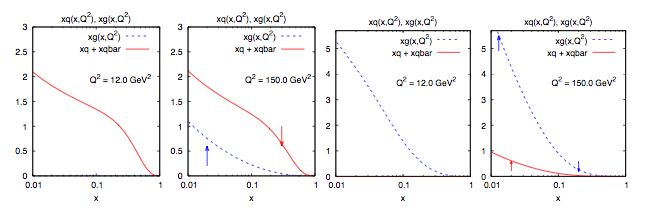}
\caption{DGLAP evolution. (left) State of quarks and anti-quarks only at $Q^2=12$ GeV$^2$ evolves into a state of quark-gluon mixture as $Q^2$ increases to 150 GeV$^2$ (middle left). (middle right) Purely gluonic state evolves into quark-gluon mixture at higher momentum scales (right). Figure provided by Ref. \cite{Salam}}
\label{fig:DGLAP}
\end{center}
\end{figure}

Allowing the PDFs to absorb soft particles suggest the possibility of renormalization. The result is the DGLAP equation \cite{Lipatov,Dok,Altarelli}. The equation describes the evolution of the PDF as a function of the factorization scale:

\begin{equation}
\frac{\mathrm{d}q(x,\mu_F^2)}{\mathrm{d}\ln \mu_F^2} = \frac{\alpha_s}{2\pi}\int_x^1\mathrm{d}zP_{qq}(z)\frac{q(x/z,\mu_F^2)}{z}, \qquad P_{qq}=C_F\left(\frac{1+z^2}{1-z}\right)_+
\label{DGLAPqq}
\end{equation}
where $P_{qq}$ is the real part of the 'Altarelli-Parisi splitting kernel' (i.e. a measure of the probability) for a quark to radiate a soft gluon. The $+$ subscript signals that the virtual part is absorbed by the expression such that divergence cancels as $z\rightarrow1$. Eq. (\ref{DGLAPqq}) describes how, when the factorization scale is increased, extra partons with momentum fraction $x$ are observed. These come from the branching of partons at lower factorization scales but with higher momentum fractions $x/z$ (Fig. \ref{fig:DGLAP}). \\
\indent The incoming proton consists not only of quarks but of gluons as well. For the complete picture all possible evolutions must be coupled into the DGLAP equation:

\begin{equation}
\frac{\mathrm{d}}{\mathrm{d}\ln\mu_F^2} \binom{q}{g} = \frac{\alpha_s(\mu_F^2)}{2\pi} 
\begin{pmatrix}
P_{qq} & P_{qg} \\
P_{gq} & P_{gg}
\end{pmatrix}
\otimes \binom{q}{g}
\label{DGLAP}
\end{equation}
where the $\otimes$ symbol is a shorthand notation for the integral in eq. (\ref{DGLAPqq}). \\
\indent In this discussion only $P_{qq}$ and $P_{gg}$ will be considered, describing the emission of a soft gluon from a quark or another gluon respectively. It turns out that in the soft and collinear limits the two splitting kernels become equivalently dependent on $z$. It will thus be convenient to assume a common kernel for the two processes of the form

\begin{equation}
P_{gg}^{qq}=\binom{C_F}{C_A}\left(\frac{2}{1-z}\right)
\label{kernel}
\end{equation}
It is worth noting that the coefficients $C_F=\frac{4}{3}$ and $C_A=3$ reveals that gluons radiate other gluons with roughly twice the probability of quarks.

\subsection{Jets}\label{sec:Jets}
Jets are crucial to collider physics since they pose an important test of perturbative QCD. Normally, the fragmentation process of parton $b$, not explicitly shown in Fig. \ref{fig:crosssection}, has to be considered. This is governed by long distance dynamics and is described by the DGLAP equations \cite{book2}, analogously to the PDFs. Thus jets become a bridge between the measurable particles in the detector and the partonic subprocess. In particular, while the latter hard process takes place on short timescales and is usually assumed not to be modified by the presence of a deconfined medium, jets traversing a hot plasma are expected to be considerably affected. 
\subsubsection{Jet fragmentaiton}\label{sec:production}
\begin{figure}
\begin{center}
\includegraphics[width=3in]{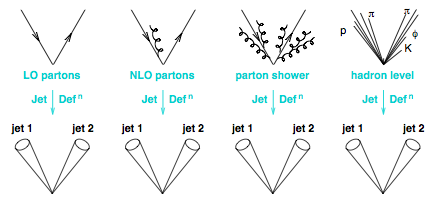}
\caption{The jet must not be sensitive to soft and/or collinear emissions, or hadronization effects. Figure provided by Ref. \cite{Salam}}
\label{fig:jetproduction}
\end{center}
\end{figure}

Returning to the simplest of situations in Fig. \ref{fig:crosssection} (section \ref{sec:crosssections}) the event $q\bar{q}\rightarrow ab$ may, for the purpose of argument, be interpreted as a $q\bar{q}\rightarrow q\bar{q}$ type event. The (anti)quarks are hard so gluon emission is expected. From eq. (\ref{emission}) the gluon is likely to be soft and close in angle to the emitter. Certainly the gluon may split into further partons and so on, and if sufficient energy is available the situation may turn into a \emph{parton shower} where each succession carries a smaller and smaller fraction of the available energy. As it seems one could end up with an indefinite spreading of particles from the initially so simple set up. However, it turns out that a parton emitted at some stage in a parton shower must always be of smaller angle than partons in a previous succession, thus resulting in increasingly soft and collinear emissions \cite{book1}. This process is referred to as \emph{angular ordering}. Rather than considering these particles individually it is convenient to cluster them into cone shaped jets of some angular resolution. The jets should not be sensitive to the softest and most collinear emissions but remain unaffected by such happenings. The process is shown in Fig. \ref{fig:jetproduction}. Notice that if a hard gluon is emitted it may well result in a jet of its own. 

\subsubsection{Hadronization - the Lund model}\label{sec:Lundmodel}
In nature, quarks are not seen as free states but are confined into hadrons. The process of hadronization is not trivial to calculate perturbatively as the hadron production must indeed involve a very large number of gluon interactions, which due to their softness is not suitable for a perturbative approach (\emph{cf} eq. (\ref{perturbation}) and eq. (\ref{coupling}), section \ref{sec:running}). However a phenomenological model (the Lund model) reproducing the main features of hadronization has been put forward \cite{Lund}. \\
\begin{figure}
\centering
\subfloat[Color field lines collapse]{\label{fig:Collapse}\includegraphics[width=0.4\textwidth]{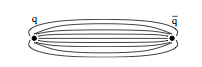}} \qquad \qquad
\subfloat[Hadronization breaks the flux tube]{\label{fig:hadronization}\includegraphics[width=0.2\textwidth, height=4cm]{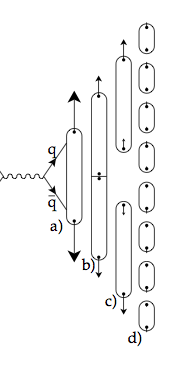}}
\caption{The process of jet formation according to the Lund model. Figure provided by Ref. \cite{Gavin}}
\label{fig:Lund}
\end{figure}
\indent Similar to electromagnetic charges, QCD color charges have associated field lines between all (anti)quarks and gluons. These lines tend to collapse into a 'string' of strong color field, with a fixed radius $\Delta r \sim \mathcal{O}(1)$ fm, hence storing a roughly constant amount of energy per unit length. Thus, in the event of quark separation a long flux tube is produced. If a $q\bar{q}$ pair is produced within the tube and therefore removing a section, then the remaining energy stored within the tube is reduced. \\
\indent Due to the force along the flux tube the hadrons will easily pick up longitudinal momentum, however any transverse component remains small ($\mathcal{O} (200)$ MeV/ c). Thus jets should be insensitive to most hadronization in addition to soft and collinear emission (Fig. \ref{fig:jetproduction} (right)). \\

\begin{figure}
\begin{center}
\includegraphics[width=4in]{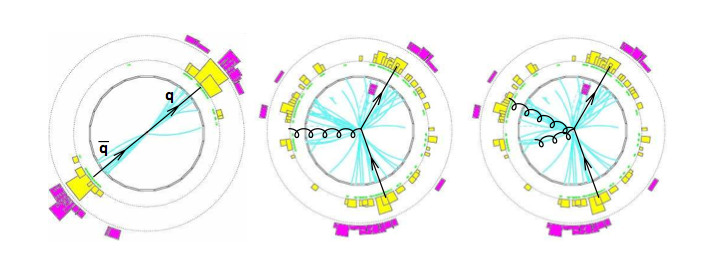}
\caption{Jet production in collider events \cite{Salam}}
\label{fig:jets}
\end{center}
\end{figure}

\subsubsection{Jet definitions}\label{sec:definitions}
\indent In some collider events it may be obvious how to define the jets from the outgoing particles such as the dijet event in Fig. \ref{fig:jets} (left). However, considering the (identical) center and right events it is not trivial to deduce whether three or four jets are observed. In order to do so a precise way of defining a jet must be provided. \\
\indent A jet definition is essentially a set of rules by which a computer can take a list of particle momenta for some event, and cluster them into jets. These rules must obey the insensitivity to soft and collinear emission. The type of definitions considered here are known as recombination algorithms. They work  in a backwards manner by intending to invert the splitting of the parton shower and recombining the particles. This does not only provide a set of jets but also the possibility of iterating the process and inspect the contents of the jet. Note that this procedure also reconstructs so-called "fake jets", i.e. random clusters of particles. \\
\indent The three recombination algorithms considered here take the form

\begin{equation}
d_{ij}(a) = \min(k_{ti}^{2a},k_{tj}^{2a}) \frac{\Delta R_{ij}^2}{R^2},  \qquad d_{iB}=k_{ti}^{2a}, 
\qquad a= 
\begin{cases}
    1 \quad \text{$k_t$-algorithm}\\
    0 \quad \text{Cambridge/Aachen}\\
    -1 \quad \text{anti-$k_t$}
    \end{cases}
\label{jetdef}
\end{equation}
The idea behind such a set up is to find a measure for the distance, $d_{ij}$, between every pair of particles $i,j$. This is then compared with some maximum allowance called the beam distance, $d_{iB}$. If the inter-particle distance is the smaller one the particles are 'combined' into a single new particle - a pseudojet. On the other hand, if the beam distance is the smaller one, then particle (or pseudojet) $i$ is denoted a jet and removed from the list of particles. The process continues until no particles remain. In eq. (\ref{jetdef}) $k_t$ is the transverse momentum with respect to the collision axis (normally the z-axis). $R$ is the opening (half) angle of a cone within which the momentum vectors of the particles in the jet are located. $\Delta R_{ij}$ is the distance between particles $i$ and $j$ in $\phi-\eta$ space such that $\Delta R_{ij}^2 = (\phi_i - \phi_j)^2 + (\eta_1 - \eta_2)^2$, where $\phi$ is the azimuthal angle of the particle and $\eta$ is the pseudorapidity defined as $\eta_i = -\ln\left[\tan\left(\frac{\theta}{2}\right)\right]$ with $\theta$ being the polar angle between the collision axis and the particle momentum. \\
\indent The three algorithms differ in the order they cluster the particles. The $k_t$ algorithm starts with the softest particles whereas the anti-$k_t$ algorithm begins with the hardest. The latter ensures that the jets are centered around a hard 'core particle', evolving in concentric circles, thus forming jet cones of well defined shape. The Cambridge/Aachen algorithm does not consider the transverse momentum at all but relies exclusively on the angular separation of the particles.

\section{Jets in a vacuum}\label{sec:vacuum}
With the theory in place it is now time to start the implementation. The objective of this discussion, as previously stated, is to test, and further develop, models suggested to replicate data obtained for hard processes taking place inside a QCD medium. However, before any QGP investigations are carried out it is essential to analyze particle behavior in a vacuum. This will be done using proton-proton collisions at a center of mass (CM) energy of 2.76 TeV. This energy is chosen as it corresponds to the CM energy of heavy-ion collisions at the LHC performed in 2010. The investigations are carried out using event generator PYTHIA 8 \cite{Sjostrand}, which we have used to generate events with a hard transverse momentum scale $\hat{p}_t=70$ GeV/c. Subsequently, the final state particles are reconstructed using the FastJet package (2.4.2) \cite{Cacciari}, which implements the recombination algorithms mentioned in section \ref{sec:definitions}. A minimum cut of 20 GeV is placed on all measured jets in order to exclude fake jets. \\
\indent PYTHIA 8 is a full fledged event generator, incorporating as one of its key features a Monte Carlo parton shower algorithm. Using the Altareli-Parisi splitting kernel and the physical phase space, see eq. (\ref{emission}), it calculates the probability of a gluon emission with given $z$ and $k_t$ and, thus, can generate a full cascade. The process continues until the ultimate gluons are below some cutoff scale where perturbation is no longer valid. This produces a parton shower. Realistic events surely consists of hadrons rather than partons, thus hadronization is considered by PYTHIA 8. This is implemented using the previously outlined Lund model.\\
\indent FastJet allows for an appropriate comparison between the three relevant recombination algorithms in order to select one for further use. A useful investigation to carry out is that of \emph{intrajet distributions}, i.e. how does the number of particles, $N$, within the jet, with a specific value of a certain observable vary as the value of the observable changes? In the following, only the hardest jet in the event is considered, and the constituent particles are measured. The variables investigated are: 

\begin{itemize}
\item $\xi=-\log\left(\frac{p_{particle}}{p_{jet}}\right)$, where $\xi$ measures the fraction of jet momentum $p_{jet}$ contained in the constituent particle,
\item $\theta=\cos^{-1} \left[\frac{\vec{p}_{particle} \cdot \vec{p}_{jet}}{p_{particle}p_{jet}}\right] $, gives the angle between the particle and the main jet, and
\item $p_t=p_{particle}\sin\theta$, which gives the particle momentum component perpendicular to the main jet, here referred to as the transverse momentum (note the difference to $k_t$ which in this discussion is used in reference to the beam axis).
\end{itemize}

\begin{figure}
\begin{center}
\includegraphics[width=6in]{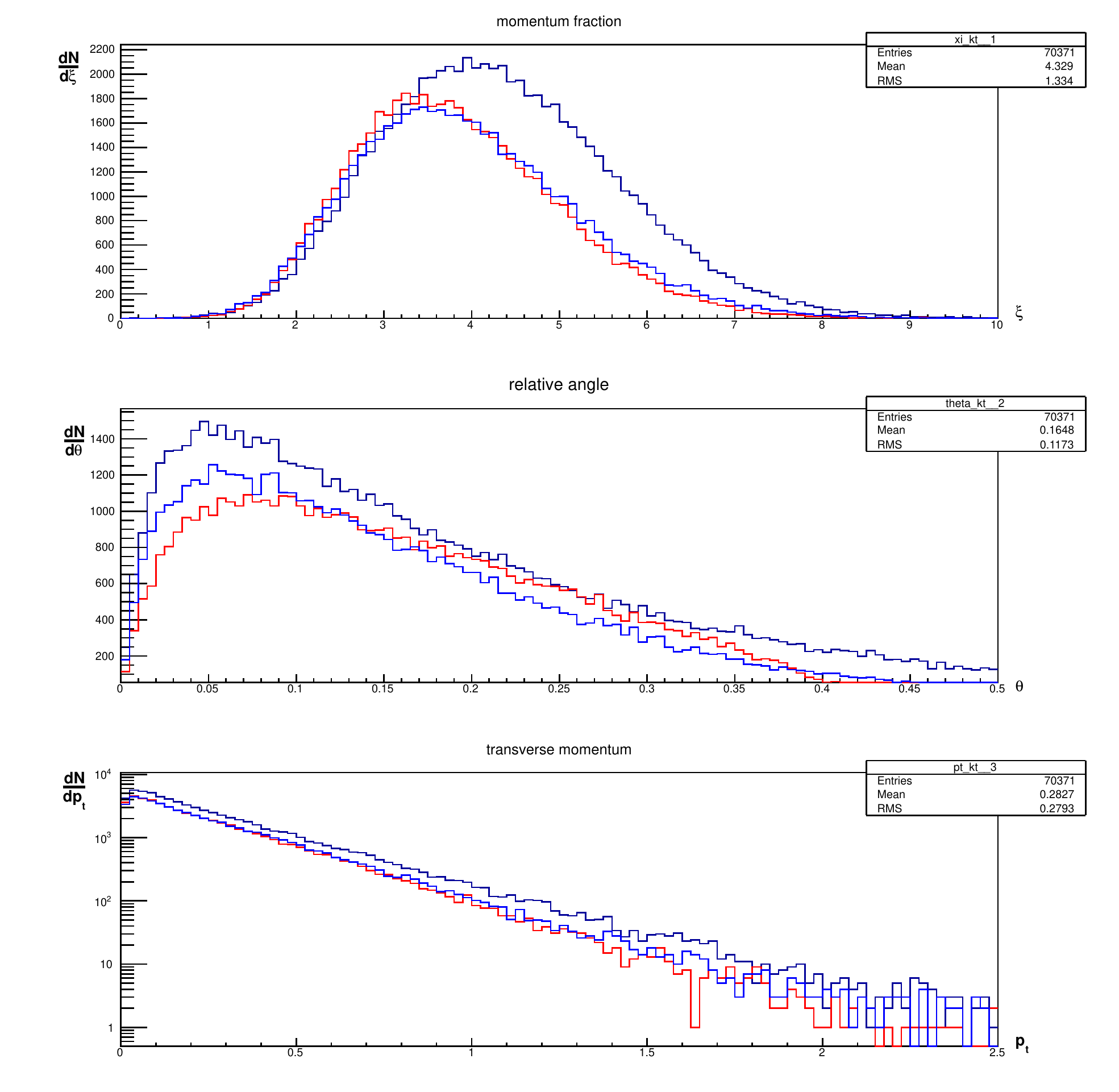}
\caption{Intrajet distributions in vacuum for the $k_t$ (black), C/A (blue) and anti-$k_t$ (red) recombination algorithms. $R=0.4$.}
\label{fig:ffvac}
\end{center}
\end{figure}

The fragmentation functions are illustrated in Fig. \ref{fig:ffvac}. A few things can be said about the jet as a whole, and the different jet definitions. Starting with $\xi$ it is evident that the distribution forms a hump and is hence referred to as the hump-backed plateau. It presents a mean of  $\sim4.3$, corresponding to a momentum fraction of $\sim 1.4\%$. It thus appears as if the jet is composed of a collection of soft particles rather than a few hard ones. This is a well known distribution that has been theoretically predicted from perturbative QCD and is a signal of angular ordering \cite{Dok2, Dok3, Bassetto}. The three algorithms agree fairly well for $\xi<3.4$ where the C/A and anti-$k_t$ algorithms peak. This corresponds to a larger momentum fraction of $\sim 4.0\%$. It appears that the $k_t$ algorithm has a large sensitivity to soft particles. This is no surprise as the algorithm uses a 'bottom up' type of approach, clustering the jet constituents starting from the softest particles to the hardest ones. This leads to jets without well defined shape - a feature which becomes even more evident when looking at the distribution over $\theta$. Despite defining a 'cone' of angle $R$ within which the particles should be, the $k_t$ algorithm ends up with larger angles than the allowed $R=0.4$. The $k_t$ 'cone' thus becomes larger and every jet contains more soft particles, which also explains the larger values for the three distributions. It is worth mentioning that the number of jet constituents with $\xi<2$ and $\xi>6$ corresponding to momentum fractions $>12\%$ and $<0.3\%$ respectively are very rare, underlining the fact that very soft and very hard jet constituents are not common. \\
\indent The $\theta$-distribution demonstrates that the anti-$k_t$ algorithm alone provides jets of a regular cone shape. This is clear from the $\theta$-axis intercept at precisely $0.4$ for  anti-$k_t$ but larger values for the other two algorithms. This is a useful trait as it allows precise geometrical limits to be put on the jets, hence this algorithm is selected for further use in this project. Non-regular jet boundaries has practical implications, which is often held against the $k_t$ and C/A algorithms. An example is in the case when an area of a detector is misbehaving. It is not trivial to see how far a $k_t$ jet must be from that area not to be influenced by it. This and other reasons have led to the anti-$k_t$ algorithm being adopted as the default jet algorithm at both ATLAS and CMS. \\
\indent It appears that the jet constituents primarily have momenta in a direction close to that of the jet as seen by the peaks of the $\theta$-distributions occurring at $0.05-0.10$ radians. This is as expected from the general collimated structure and formation of jets as was discussed in section \ref{sec:production}. \\
\indent The $p_t$-distribution indicates again that the majority of jet constituents carry a very small momentum compared to the jet as a whole. The distribution appears to obey an exponential law up until values of $p_t \sim 2$ GeV/c after which the number of recorded readings become very small and the statistics unreliable. \\

A further insight into the distribution of energy within the jet can be done by defining a transverse momentum density $\rho_\perp(r )$ at some angle $r<R$ relative to the axis of the jet. Then the fraction of transverse jet momentum relative to the beam axis, contained within the 'subcone' of opening angle $r$ is

\begin{equation}
\psi(r )=\frac{\int_0^r \rho_\perp(r')dr'}{p_t^{jet}}
\end{equation}

\begin{figure}[h]
\begin{center}
\includegraphics[width=3.9in]{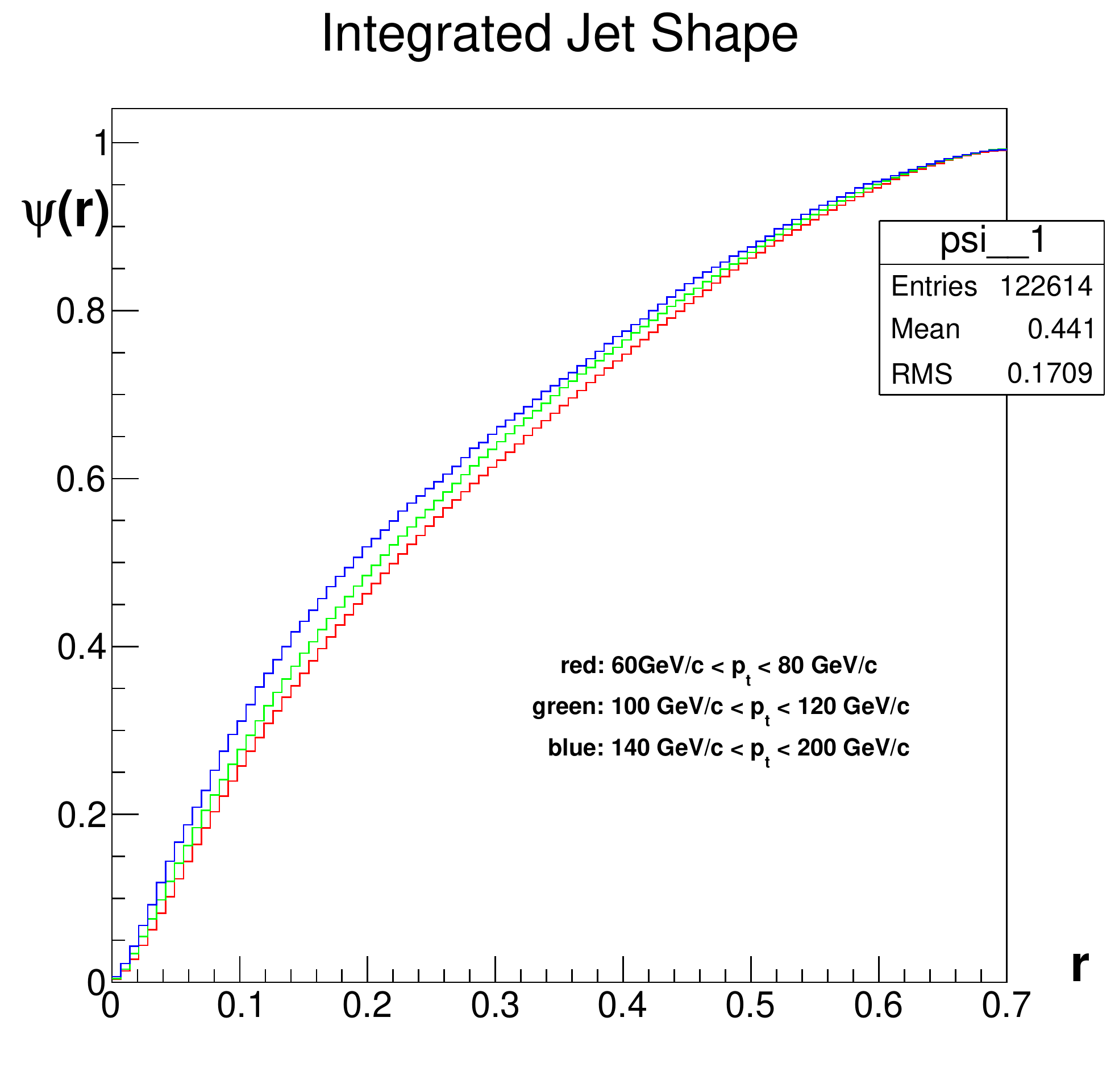}
\caption{$\sim 75\%$ of the jet energy is concentrated in the innermost half of the jet. More energetic jets are more collimated than less energetic ones.}
\label{fig:jetshape}
\end{center}
\end{figure}

\begin{figure}[t]
\begin{center}
\includegraphics[width=3.9in]{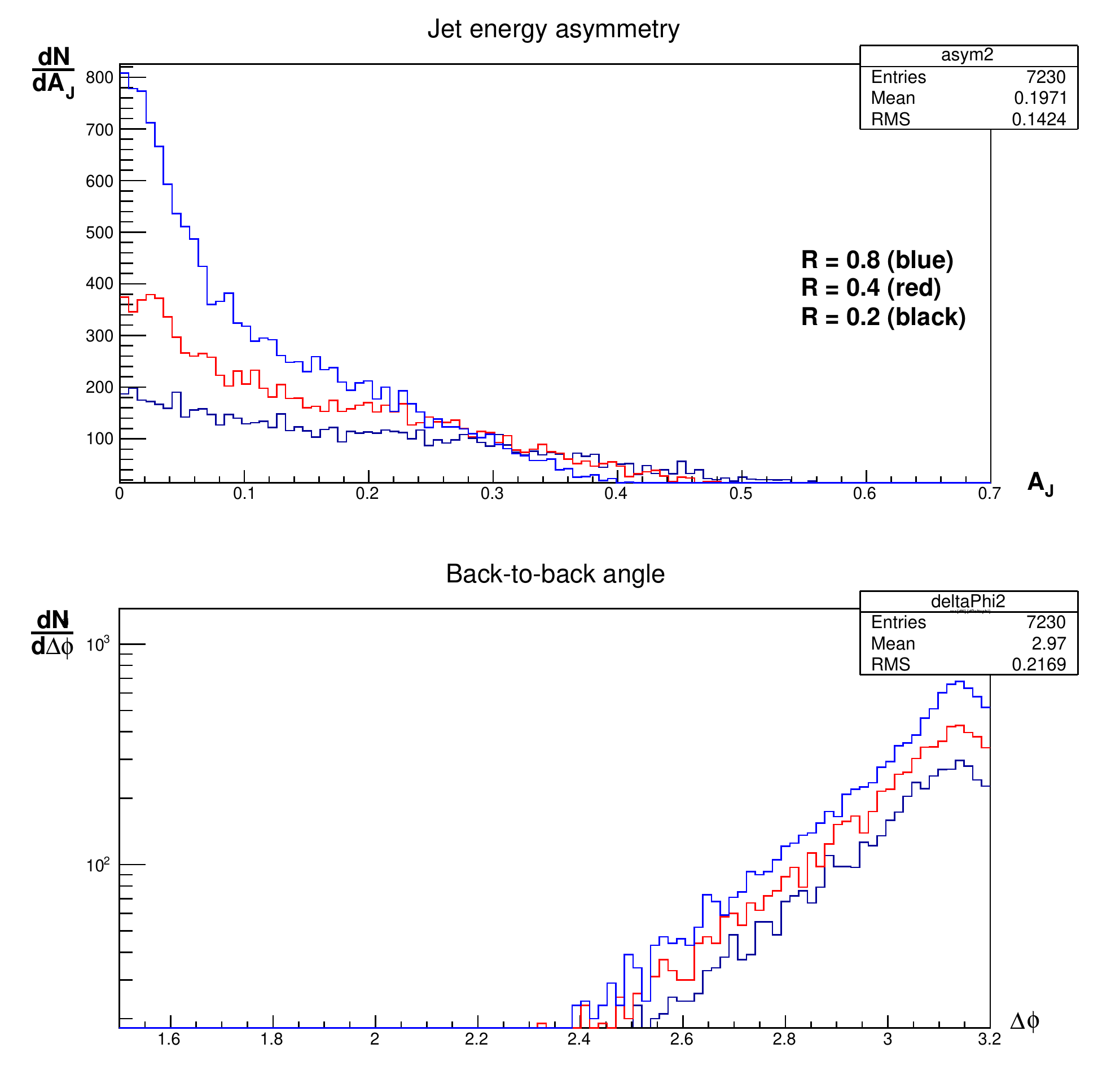}
\caption{Asymmetry and back-to-back angle distributions in vacuum for anti-$k_t$ algorithm. $p_t^{min}=20$ GeV/c.}
\label{fig:asymvac}
\end{center}
\end{figure}

Fig. \ref{fig:jetshape} illustrates what fraction of the jet energy is contained within the subcone. The jets appear rather collimated with approximately 75\% of the energy concentrated in the innermost half of the jet and the most central third carrying about 50\% ($R=0.7$). It is worth noting that the more energetic jets appear more collimated (blue) than the less energetic ones (red). This means that despite increased radiation, the harder jets still carry most energy in the central core particles of the jet. \\

It is of interest to study how the jets resulting from a collision balance each other. A convenient way of doing this is to measure the jet energy asymmetry, $A_J$, and the azimuthal angle separation between the two hardest jets (Fig. \ref{fig:asymvac}), defined as

\begin{equation}
A_J=\frac{E_{T1}-E_{T2}}{E_{T1}+E_{T2}}, \qquad |\Delta\phi|=|\phi_1-\phi_2|.
\end{equation}
Here $E_{T1}$ and $E_{T2}$ are the transverse energies of the hardest and second to hardest jets respectively, and $|\Delta\phi|$ is the angle between them. Thus, an asymmetry close to zero indicates that a collision has resulted in two jets of approximately equal energies. A value strongly deviating from zero indicates that more than two jets are present. \\
\indent Here a limit have been put on the jet transverse momentum in order to avoid contributions from background from the underlying event. For the hardest jet $E_{T1}>100$ GeV and for the second jet $E_{T2}>25$ GeV. Additionally, we demand that both of the jets are at mid-rapidity ($|\eta| <1$). The asymmetry distribution in Fig. \ref{fig:asymvac} suggests that a large cone results in many jets whereas a small one results in a few. This is largely a result of the cutoffs described above, introduced to reduce the contribution from soft background partons. Thus, some jets, clustered with a small value for the parameter $R$, are not centered around a hard core particle (such as may be the case in Fig. \ref{fig:jets} (right), section \ref{sec:definitions}) and will be eliminated. \\
\indent It appears that if a large jet cone is used, then most events will result in two jets whereas for very small cones the number of jets per event increases, as is seen from the progressively shallower slopes at lower opening angles. This is certainly expected due to the sheer geometry of the problem. It is worth noting is that a very large cone may engulf several particle clusters into one jet, thus miss a lot of interesting physics. It will be convenient for further reference to limit the cone size to $0.4<R<0.7$. \\
\indent The $|\Delta\phi|$-distribution indicates that most often the two hardest jets are separated by an angle close to $\pi$ radians, again supporting the idea that most events result in two major jets, possibly accompanied by one or a few soft ones. \\

\section{Jets in a QCD medium}\label{sec:QGP}
Experiments at RHIC have established a strong suppression of hadrons with high transverse momentum in hard heavy-ion processes \cite{PHENIX,STAR}. This suggests that there is an interaction between high $p_t$ partons and QCD matter resulting in a significant loss of energy for the partons in addition to the vacuum fragmentation - a process referred to as 'jet-quenching'. It has been suggested that the underlying reason is medium-induced gluon radiation \cite{Baier}. \\
\indent The introduction of a background medium requires the mathematical models of QCD to be modified to account for two main effects: Primarily all hard particles must be prompted to radiate gluons at a higher rate than in vacuum. Secondly, the increased gluon emission must cause a momentum spread among the particles. \\
\begin{figure}
\begin{center}
\includegraphics[width=4in]{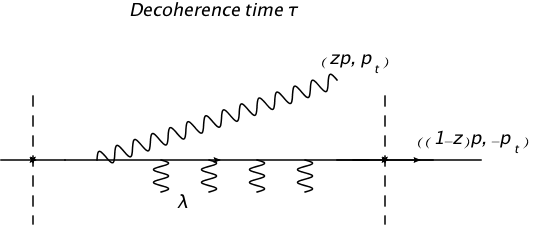}
\caption{In a QGP medium induced gluon emission should result in additional soft partons on the expense of hard hadrons. During the formation time $\tau$ needed for the gluon to decohere from the initial parton shower a typical number of collisions $N_{coh}=\frac{\tau}{\lambda}$ occurs giving the gluon a transverse 'kick'. }
\label{fig:collimation}
\end{center}
\end{figure}
\indent A suggested model is that the partonic scattering cross sections are dominated by small-angle scattering. Multiple small-angle scatterings has the consequence that all components of the parton shower undergo Brownian motion, a trait that can be used to quantify the spreading of the jet energy \cite{Casalderrey}. \\
\indent Fig. \ref{fig:collimation} pictures the emission of a gluon from some hard particle during the formation time $\tau$. It is straight forward to calculate the momentum uncertainty, hence using Heisenberg's uncertainty principle (HUP) find an expression for $\tau$:

\begin{equation}
\begin{aligned}
&\Delta p = \sqrt{(1-z)^2p^2 + p_t^2} + \sqrt{z^2p^2 + p_t^2} - p \sim \frac{p_t^2}{p}\\
&\Delta t\Delta p = \tau\Delta p \sim 1 \qquad \Rightarrow \qquad \tau \sim \frac{p}{p_t^2} = \frac{\omega}{p_t^2}
\label{eq:HUP}
\end{aligned}
\end{equation}
where $\omega$ is the energy of the hard particle. Notice that in a vacuum $p_t=\omega\sin{\theta}\simeq\omega\theta$ meaning that $\tau\propto\frac{1}{\omega}$ for a fixed angle. Thus in a vacuum a soft particle takes longer to decohere from the parton shower than a hard one does. \\
\indent Returning to Fig. \ref{fig:collimation}, during the gluon formation time, in a medium the particle may undergo a series of collisions. The typical number of collisions can be expressed as $N_{coh}=\frac{\tau}{\lambda}$ where $\lambda$ is the mean free path of the particle. If on average each collision contributes with transverse 'kick' $\mu$ the average transverse momentum, $\langle p_t^2 \rangle=\mu^2N_{coh}=\frac{\mu^2}{\lambda}\frac{\omega}{p_t^2}$. It is convenient to denote by $\hat{q}$ the average squared transverse momentum per unit path length, that the gluon accumulates in the medium. Thus;

\begin{equation}
p_t^2=\sqrt{\hat{q}\omega}, \qquad \tau=\sqrt{\frac{\omega}{\hat{q}}},
\label{eq:taumed}
\end{equation}
from which it is evident that in a medium the soft particles decohere faster than the hard ones, i.e. contrary to the behavior in a vacuum. Thus, in this model of a QGP softer gluons decohere faster from the parton shower due to small-angle scattering induced by the medium. \\
\indent The intensity of the gluon emission across the medium of size $L$ can be expressed as

\begin{equation}
\begin{aligned}
&dI\propto \frac{\alpha_s}{2\pi}\binom{C_F}{C_A}\frac{d\omega}{\omega}\frac{L}{\lambda N_{coh}}=\\
&= \frac{\alpha_s}{2\pi}\binom{C_F}{C_A}\frac{d\omega}{\omega}\frac{\mu^2}{\lambda}\frac{L}{p_t^2}=\\
&= \frac{\alpha_s}{2\pi}\binom{C_F}{C_A}\frac{d\omega}{\omega}\frac{\hat{q}L}{\sqrt{\hat{q}\omega}}=\\
&= \frac{\alpha_s}{2\pi}\binom{C_F}{C_A}\frac{d\omega}{\omega}\sqrt{\frac{\omega_c}{\omega}}
\end{aligned}
\label{intensity}
\end{equation}
where in the final line some critical energy scale, $\omega_c$, has been identified due to the dimensionless nature of the intensity. The color factor $C_F$ ($C_A$) applies for radiation off a quark (gluon). By inspection $\omega_c=\hat{q}L^2$, leading to the observation that

\begin{equation}
\tau(\omega_c)=\sqrt{\frac{\hat{q}L^2}{\hat{q}}}=L, \qquad p_t^2(\omega_c)=\hat{q}L
\label{transmom}
\end{equation}
i.e. the formation time for the gluon is the entire size of the quark-gluon plasma, meaning that the critical energy previously identified in fact is the maximum energy the gluon is allowed to receive from the medium. Thus an upper limit is introduced to the momentum gain of the partons, hence also a limit on the momentum spread. It is clear that for soft gluons of energy $\omega \le \sqrt{\hat{q}L}$ will be completely decorrelated from the initial jet direction, thus the observation is consistent with the idea that soft particles decohere more rapidly than hard ones. \linebreak

\subsection{Toy model for qualitative observations}
Based on the above considerations, we proceed with a simple implementation of two main medium modifications of the jet fragmentation that we hope will capture the main physical mechanisms involved. Firstly, the spectrum in eq. (\ref{intensity}) suggests an enhancement of soft gluon production due to interactions with the medium. Secondly, soft gluons accumulate transverse momentum along their path through the medium, see eq. (\ref{transmom}), resulting in their Brownian motion. This, in turn, will lead to a decollimation of the soft constituents of a given jet. \\
\indent Our approach to implementing the former of the above outlined effects is a modification of the Altarelli-Parisi splitting kernels (eq. (\ref{kernel}), section \ref{sec:DGLAP}), suggested by Borghini and Wiedemann \cite{Borghini}. The idea behind this modification is to significantly increase the likelihood for soft gluon emission by enhancing the soft part of the splitting function as follows

\begin{equation}
P_{gg}^{qq} = \binom{C_F}{C_A} \frac{2+f_{med}}{1-z}
\end{equation}
where $f_{med}$ is a constant. At this stage in the investigation it is merely of interest to observe what qualitative effects the modification may have. Therefore, $f_{med}$ is arbitrarily set to equal one. Borghini and Wiedemann \cite{Borghini} suggests that the modification should be constant, $f_{med}=0.8$. This value is chosen as it has allowed the aforementioned to replicate the suppression of the $\pi^0$ meson in Au-Au collisions at $\sqrt{s}=200$ GeV, see Fig. \ref{fig:RAA}. \\
\indent An idea presented by Casalderrey-Solana, Milhano and Wiedemann \cite{Casalderrey} is to assign a 'kick' to all particles traversing the quark-gluon plasma, a concept referred to as "jet collimation". We assume that the induced gluon emission also takes place purely perpendicular to the beam axis. Every final state hadron is assigned an additional momentum kick of the order $\sqrt{\hat{q}L}$ perpendicular to their trajectory, see eq. (\ref{transmom}). In our implementation, this extra boost follows a Gaussian distribution with standard deviation $\sigma=\hat{q}L$ in accordance with the above argument. For the moment, $L$ is is assumed to depend linearly on the azimuthal angle, $\phi$. Again, qualitative effects are sought for so $\hat{q}$ is arbitrarily set to 1 GeV$^2$/fm. The path length through the medium is allowed to vary according to $L=L^* + (L_o - L^*)\frac{|\phi|}{\pi}$, where $L^*$ is the maximum allowed medium size at $\phi=0$, and $L_o$ is some lower cut. Here $L^*=1.0$ fm and $L_o=0.5$ fm. Thus it is assumed that the shortest distance through the medium occurs at zero angle in azimuth.\linebreak 

\begin{figure}
\makebox[\textwidth][c]{
\centering
\subfloat[ATLAS data]{\label{fig:ATLASdata}\includegraphics[width=0.35\textwidth]{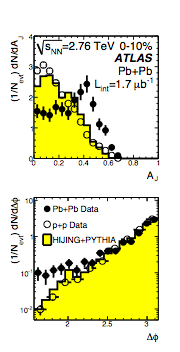}}
\subfloat[Modified splitting kernel]{\label{fig:epsmodel}\includegraphics[width=0.45\textwidth]{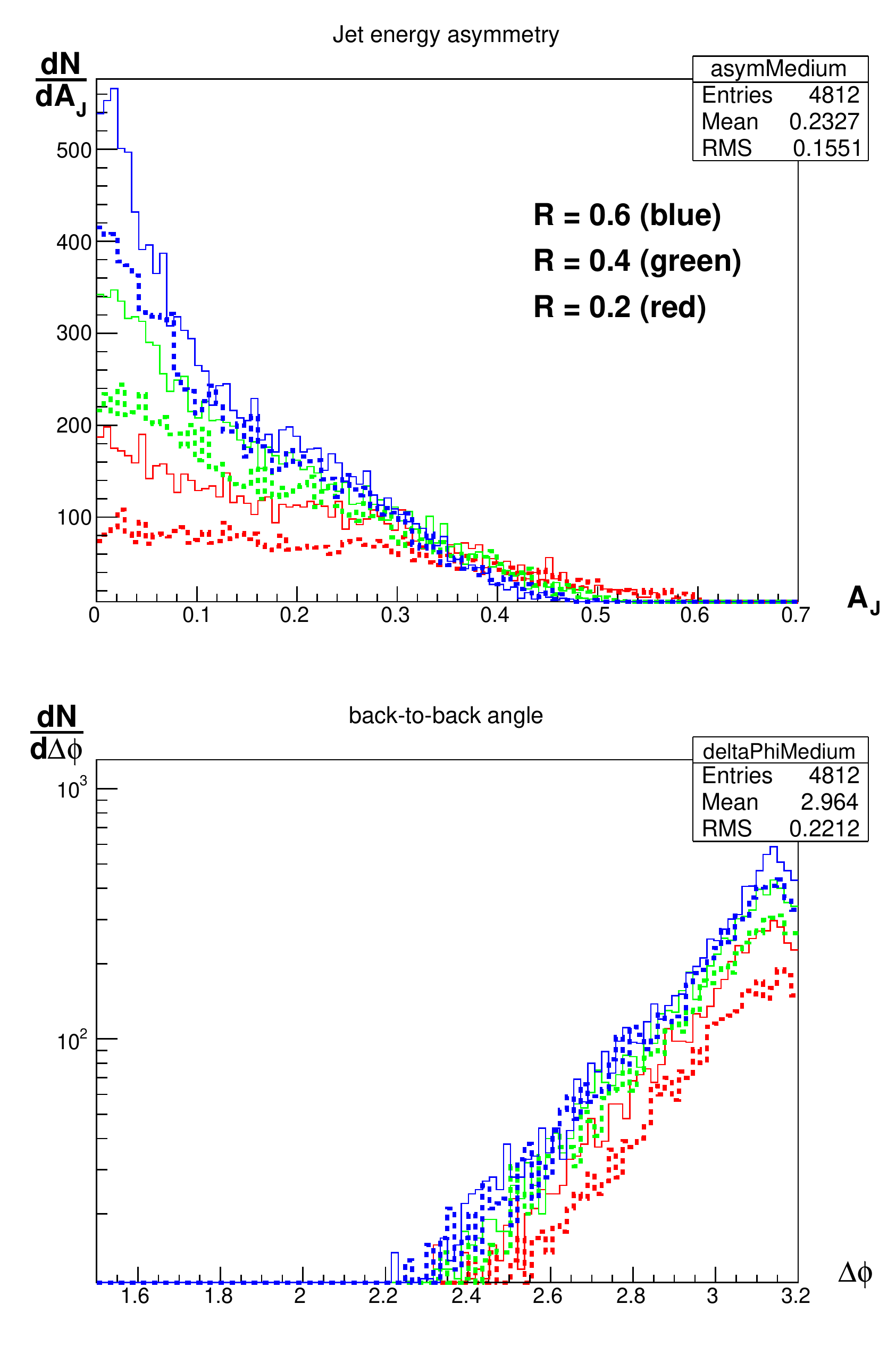}}
\subfloat[Additional jet collimation]{\label{fig:collimmodel}\includegraphics[width=0.45\textwidth]{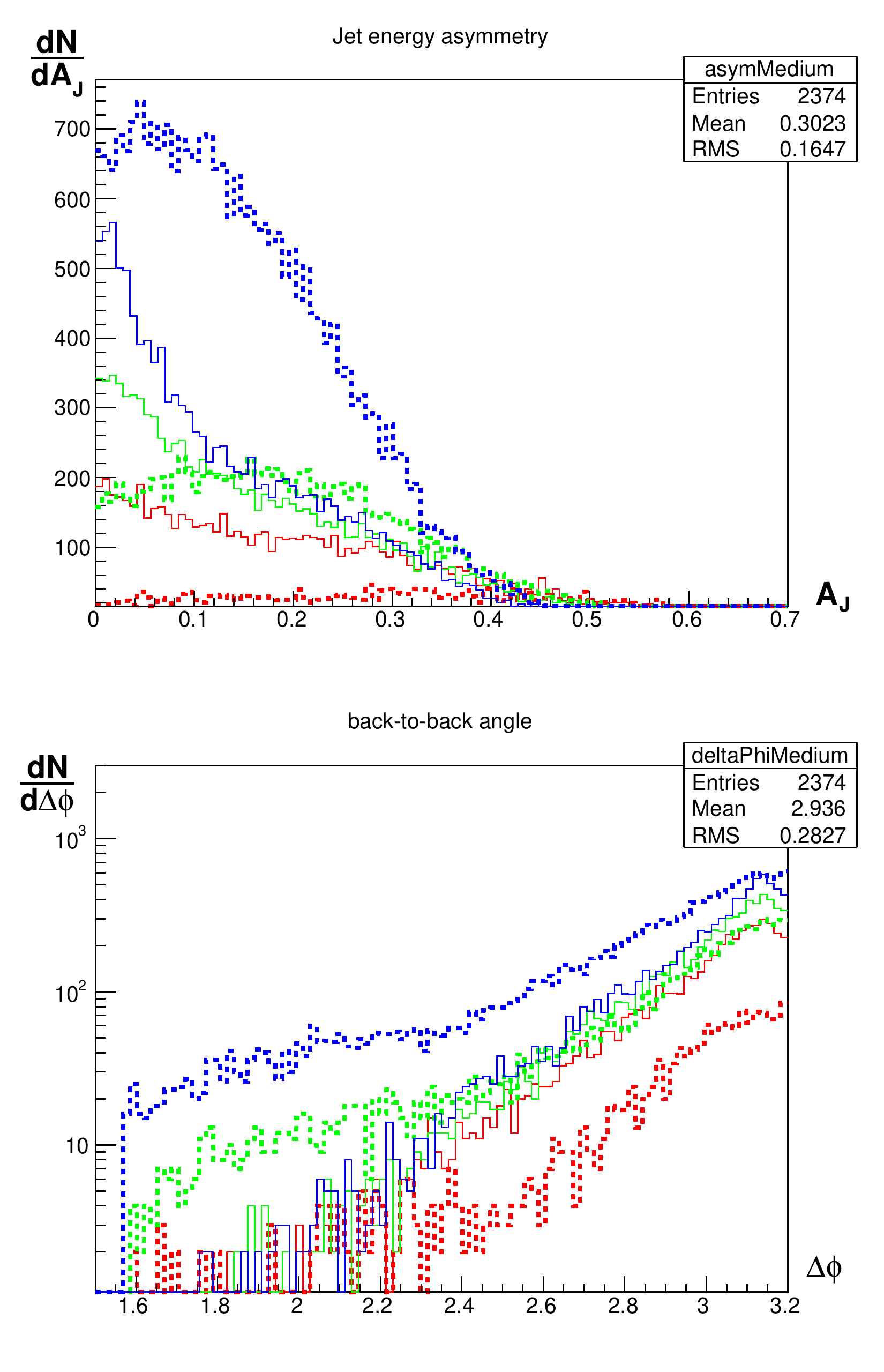}}
}
\caption{\\(a) Asymmetry and back-to-back angle of the two hardest jets as measured at ATLAS. Yellow bins are PYTHIA events for pp collisions in vacuum at $\sqrt{S} = 7$ TeV, empty circles are corresponding experimental data and solid black circles are data from Pb-Pb collisions at $\sqrt{S}=2.76$ TeV. $R=0.4$. 
\\ (b) Results from modified Altarelli-Parisi splitting kernel. Solid line shows vacuum results and dotted line shows medium results.
\\ ({c}) Results for modified splitting kernel in combination with jet collimation.
\\ $f_{med}=1$
\\$\hat{q}=1$ GeV$^2$/fm}
\label{fig:ATLAS}
\end{figure}

\begin{figure}[h]
\begin{center}
\includegraphics[width=3.9in]{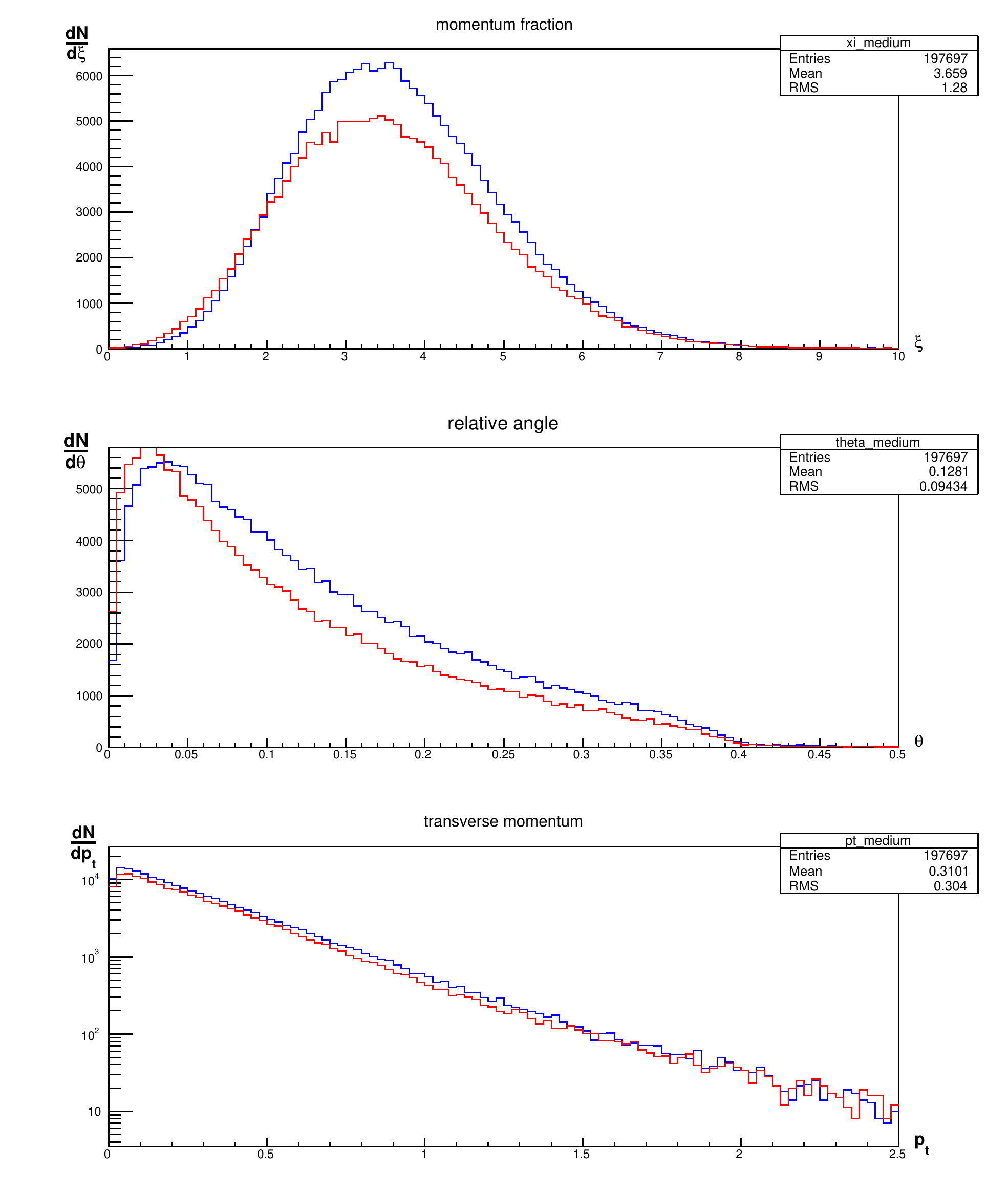}
\caption{Intrajet distributions for processes in vacuum (red) and in quark-gluon plasma (blue) ($R=0.4$).}
\label{fig:ffMedium}
\end{center}
\end{figure}

\indent The ATLAS and CMS Collaborations reported after the first $\sqrt{s}=2.76$ TeV Pb-Pb collisions in 2010 their observations of jet energy asymmetry and azimuthal angle separation between the hardest and second jet \cite{ATLAS}\cite{CMS}. The relevant results are shown in Fig. \ref{fig:ATLASdata}. As before, a limit has been put on the jet transverse momentum in order to avoid contributions from background from the underlying event, namely $E_{T1}>100$ GeV and $E_{T2}>25$ GeV. The data indicates an increased level of asymmetry for the Pb-Pb collisions compared to proton-proton collisions in vacuum. A clear broadening is seen; a shift to a higher mean; and a shift of the zero peak to a larger value. Here a cone opening angle of $R=0.4$ was used. The ATLAS Collaboration further reported that the asymmetry increased for a smaller $R$. \\
\indent It is evident from Fig. \ref{fig:epsmodel} that merely modifying the splitting kernel with a constant does not replicate the characteristics observed at ATLAS. The result indicates that this rather alters the asymmetry and azimuthal separation such that they behave as if a smaller opening angle $R$ was used. Fewer entries in the histogram suggests that the jets have less energy. This supports the proposition that gluon emission has increased and somehow transports energy out from the jet cone. However, the general behavior of the jets remain. This is increasingly obvious if one considers Fig. \ref{fig:ffMedium}. It is clear that the number of particles involved in the hard process increases as the splitting kernel is modified. The $\xi$-distribution reveals that  the number of very hard particles, i.e. those of low $\xi$, have decreased somewhat whereas the softer particles have increased in numbers. This indicates that the hard core particles of the jet are suppressed by the medium and radiating its energy as soft gluons as expected. The shift of the peak in the $\theta$-distribution suggest a slight spread of the particles, however any alteration in the $p_t$ distribution is insignificant. It may thus be concluded that the main effect of this modification is to increase to soft emissions.

Considering Fig. \ref{fig:collimmodel}, picturing the results from the collimation modification, it is clear that the behavior is dependent on the jet cone opening angle, $R$. It must be noted that the particles in the jets are assigned with additional transverse energy of $\mathcal{O}(1)$ GeV/c. For a large jet cone (blue) this additional momentum may not be enough to remove most gluons from the jet but rather spread them to the edges of the cone, thus resulting in an overall increased broadened transverse jet energy distribution. However, for a small jet cone (red) the decoherence may be quite sufficient to strip the jet of most gluons, hence reducing the over all energy. This explains the significant increase (reduction) of entries in the histogram for $R=0.6$ ($R=0.2$). For $R=0.4$ (green) these effects seem to approximately cancel and the number of entries are of the same order for the vacuum (solid) and the medium (dotted) models. Just as the ATLAS and CMS Collaborations observed \cite{ATLAS,CMS}, in Fig. \ref{fig:ATLAS} a clear broadening is seen in the asymmetry; a shift to a higher mean, here by approximately 30\%; and a shift of the zero peak to a larger value which seems to increase somewhat for smaller jet cones. The $\Delta\phi$-distribution reveals a tailing towards low values starting at $\sim2.5$ for $R=0.4$, in rough agreement with ATLAS and CMS. \\
\indent It appears that a random kick in the transverse plane \cite{Casalderrey} in combination with the modified splitting kernel \cite{Borghini} provides the correct qualitative features of the jet energy asymmetry distribution and the back-to-back behavior of the hardest and second jet. However, for a more realistic picture a discussion of the geometry of the collision, and appropriate selections of the parameters $f_{med}$ and $\hat{q}$ are in place.

\subsection{Refined model - Nuclear geometry}

\begin{figure}
\centering
\subfloat[Geometry of Pb-Pb collision]{\label{fig:geometrydrawing}\includegraphics[width=0.65\textwidth]{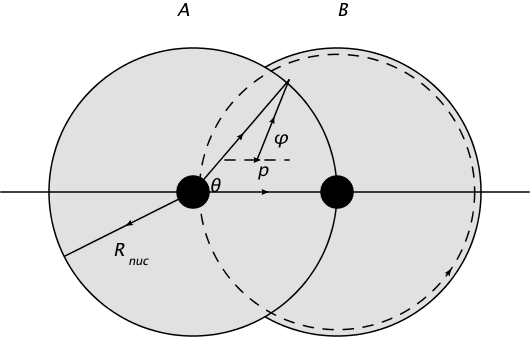}}
\subfloat[Random creation points and path length]{\label{fig:geometryplot}\includegraphics[width=0.45\textwidth]{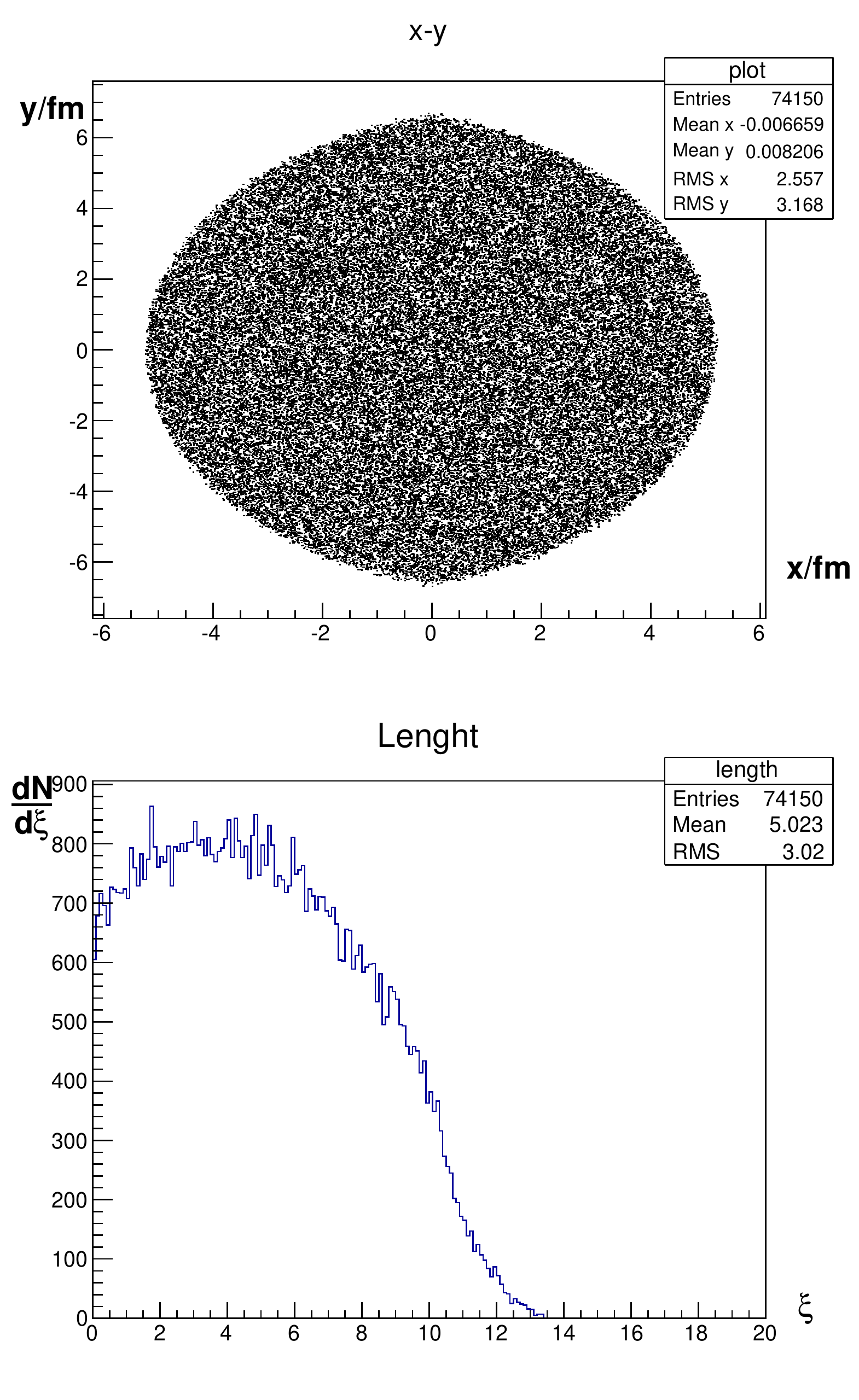}}
\caption{\\(a) Simple geometrical representation of a collision between lead nuclei, labelled A and B.
\\ (b) (top) Randomly generated points of creation for jets. (bottom) Distribution of path lengths through quark-gluon plasma.}
\label{fig:geometry}
\end{figure}

At the energies dealt with  the incoming lead nuclei will be Lorenz contracted and appear virtually two dimensional in the collision moment. In a simple model the nuclei can thus be assumed to take the shape of circular 'pancakes' of radius $R_{nuc}\simeq1.15A^{1/3}\simeq6.8$ fm, with $A=207$ being the mass number for lead. Fig. \ref{fig:geometrydrawing} illustrates how the overlap of these pancakes forms an almond shape in a semi-peripheral collision with impact parameter $b=3.2$.\footnote{The impact parameter is defined as the distance between the center of the two colliding nuclei.} For the time being it will be assumed that this shape is filled with a quark-gluon plasma from the time of the collision, and remains static and unchanged throughout all measurements. It is necessary to calculate the distance from a jet creation point p($x_o,  y_o$) to the edge of the almond. Assuming the path through the medium is a straight line and that the jet exits through a segment of a circle centered at ($x_{Ao},  y_{Ao}$) one can write;

\begin{equation}
\begin{aligned}
& x_A=x_{Ao}+R_{nuc}\cos{\theta} \qquad & x=x_o+\xi\cos{\phi}\\
& y_A=y_{Ao}+R_{nuc}\sin{\theta} \qquad & y=y_o+\xi\sin{\phi}\\
& \text{for circle A} \qquad & \text{for linear path}
\end{aligned}
\end{equation}
and similar for circle B. Equating gives

\begin{equation}
\xi=\left[ R_{nuc}^2 - (\Delta y\cos{\phi} - \Delta x\sin{\phi})^2 \right]^{1/2} - \Delta x\cos{\phi} - \Delta y\sin{\phi}
\end{equation}
where $\Delta x$ and $\Delta y$ are the coordinates of the jet production point relative to the center of the pancake through which edge the jet exits. $\xi$ is now the distance to the almond edge as a function of the jet direction in the transverse plane, the azimuthal angle $\phi$, and the point of production. Thus a randomly selected production point can be assigned to all created jets as is illustrated in fig. \ref{fig:geometryplot} (top). Here the distance between the nuclear centers is 3.2 fm corresponding to an overlap of roughly 70\%. The bottom picture indicates that the majority of particles travels less than 8 fm through the medium with a mean around 5 fm. \\
\indent Using $\xi$ as the path length $L$ it is time to return to  the previous model. There is no longer any need to assume a linear behavior of $L$ in $\phi$. $L$ varies as derived above and it is reasonable to expect that the modification of the splitting kernel could vary linearly in $L$ such that 
\begin{equation}
f_{med}(L) = Lf_{med}^* + f_{med}^o
\end{equation}
where  $f_{med}^o$ is some minimum modification (corresponding to a jet production point at the edge of the almond overlap and traveling outwards, hence set to zero) and $f_{med}^*$ is a characteristic modification. Since we have no handle on the space-time structure of the jet fragmentation, we will additionally assume that all particles originate from a point where the hard process has taken place. This calls for an improvement to be addressed in the future. \\
\indent We have tested two options: (i) a constant value $f_{med}^o=0.8$ \cite{Borghini} and $f_{med}^*=0$, and (ii) a variable $f_{med}(L)$ with parameters $f_{med}^o=0$ and $f_{med}^*=0.8/5$, due to the mean path length being close to 5 fm, see Fig. \ref{fig:geometry}. \\
\begin{figure}
\centering
\subfloat[Modified splitting kernel alone]{\label{fig:fmed}\includegraphics[width=0.5\textwidth]{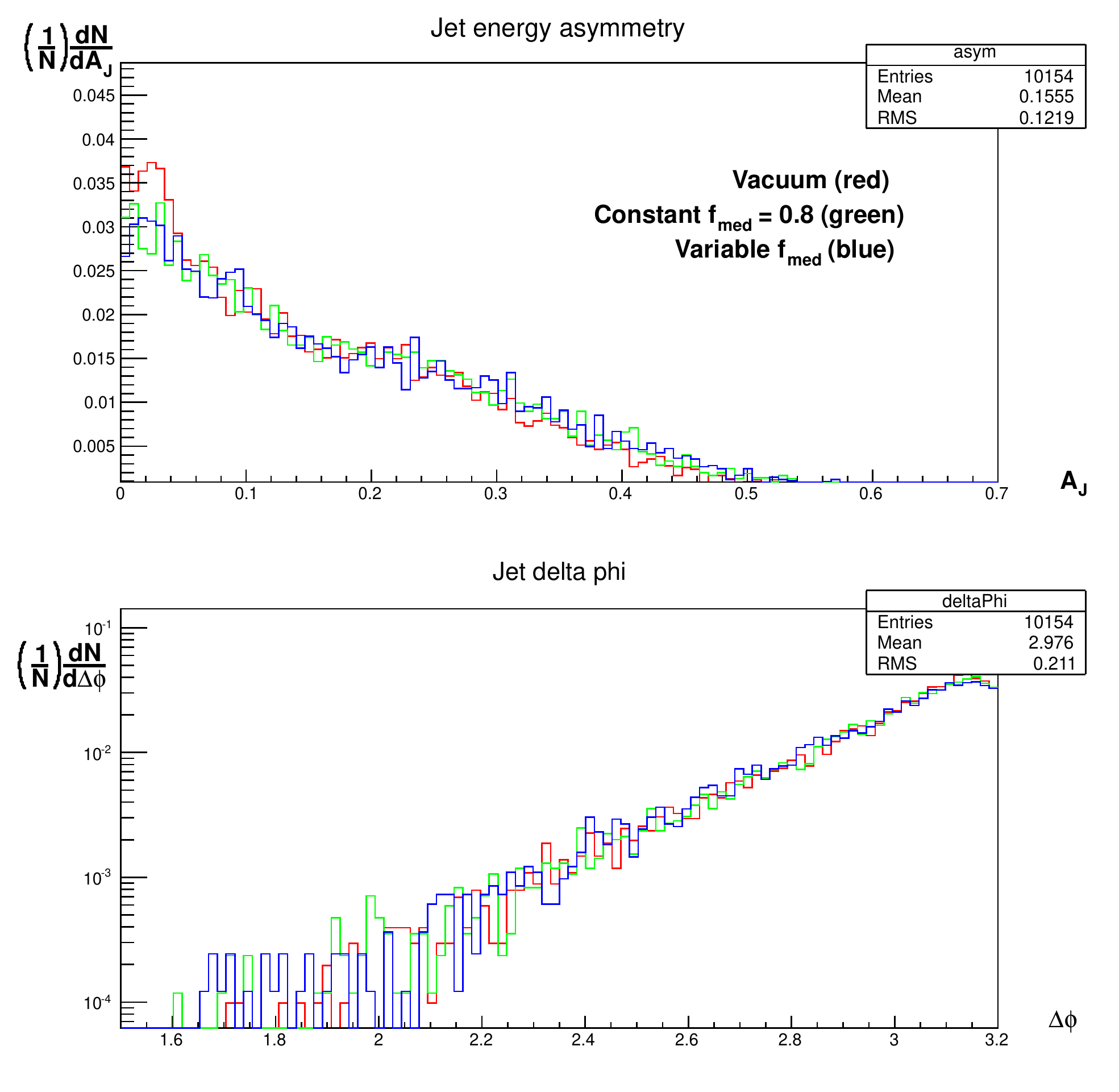}}
\subfloat[Additional jet collimation]{\label{fig:decoherence}\includegraphics[width=0.5\textwidth]{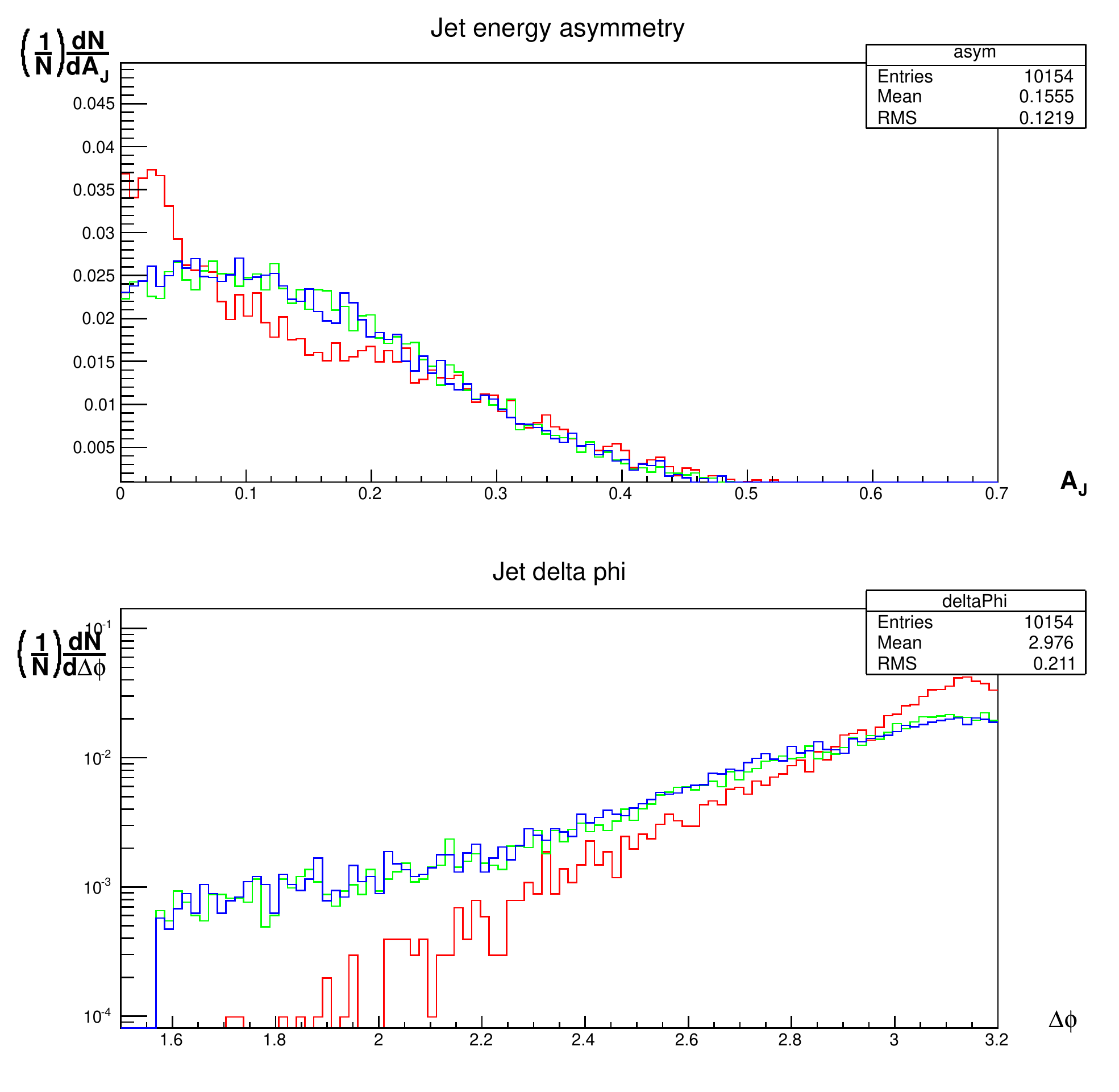}}
\caption{Asymmetry and azimuthal angle separation for refined QCD model.}
\label{fig:asymfinal}
\end{figure}
\indent Fig. \ref{fig:fmed} immediately reveals that the difference between a variable and a constant $f_{med}$ is insignificant. Here the histograms have been normalized to the number of entries to simplify the comparison with the ATLAS data (Fig. \ref{fig:ATLAS}), and $R=0.4$. It is clear that modifying the splitting kernel alone is an insufficient model to describe the changes a jet experiences as it passes through the quark-gluon plasma. Adding the jet collimation mechanism to the model replicates the ATLAS data with improved accuracy as is shown in Fig. \ref{fig:decoherence}. There is a clear broadening in the asymmetry distribution and significant increase of hard jets with azimuthal angle separation strongly deviating from $\pi$. The model does not replicate the clear peak seen in the asymmetry spectrum obtained at ATLAS. Neither does it reproduce the distinct tail observed in the $\Delta\phi$-spectrum at ATLAS but rather alters the gradient of the plot. An interesting feature is that for asymmetries above $\sim0.22$ there is no distinct difference between the vacuum and the medium situations. This may be a critical value where three jet events become increasingly frequent, as is suggested by the little hump in the vacuum distribution. There is a vague indication in the ATLAS data that the gradients at large asymmetries should be of the same order. Thus three jet events appear to be less affected by the medium than dijet events. This can of course be experimentally investigated. \\
\begin{figure}
\makebox[\textwidth][c]{
\centering
\subfloat[Angular ordering ensures additional gluon emissions collimates the jet in vacuum. ]{\label{fig:vacuumcone}\includegraphics[width=0.35\textwidth]{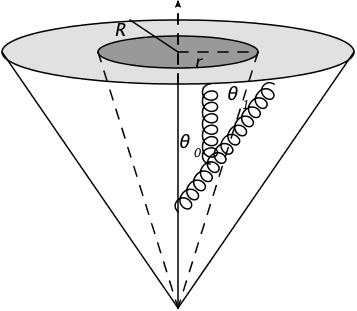}} \qquad
\subfloat[The QGP spreads gluons towards the cone edge or even outside the jet.]{\label{fig:medcone}\includegraphics[width=0.35\textwidth]{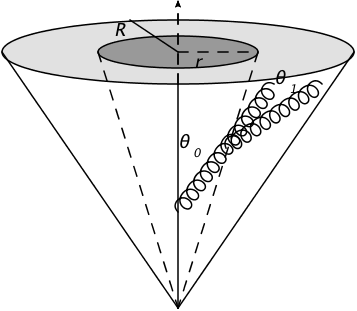}} \qquad
\subfloat[Collimation for a jets in (a) (black) and (b) (red).]{\label{fig:psi}\includegraphics[width=0.35\textwidth]{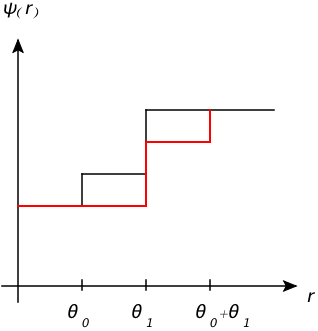}}
}
\caption{Analysis of jets consisting of a single hard particle emitting a gluon suggests that jets in a QGP should be less collimated than jets in a vacuum.}
\label{fig:collimtheo}
\end{figure}
\begin{figure}
\makebox[\textwidth][c]{
\centering
\subfloat[$R=0.2$]{\label{fig:collim2}\includegraphics[width=0.35\textwidth]{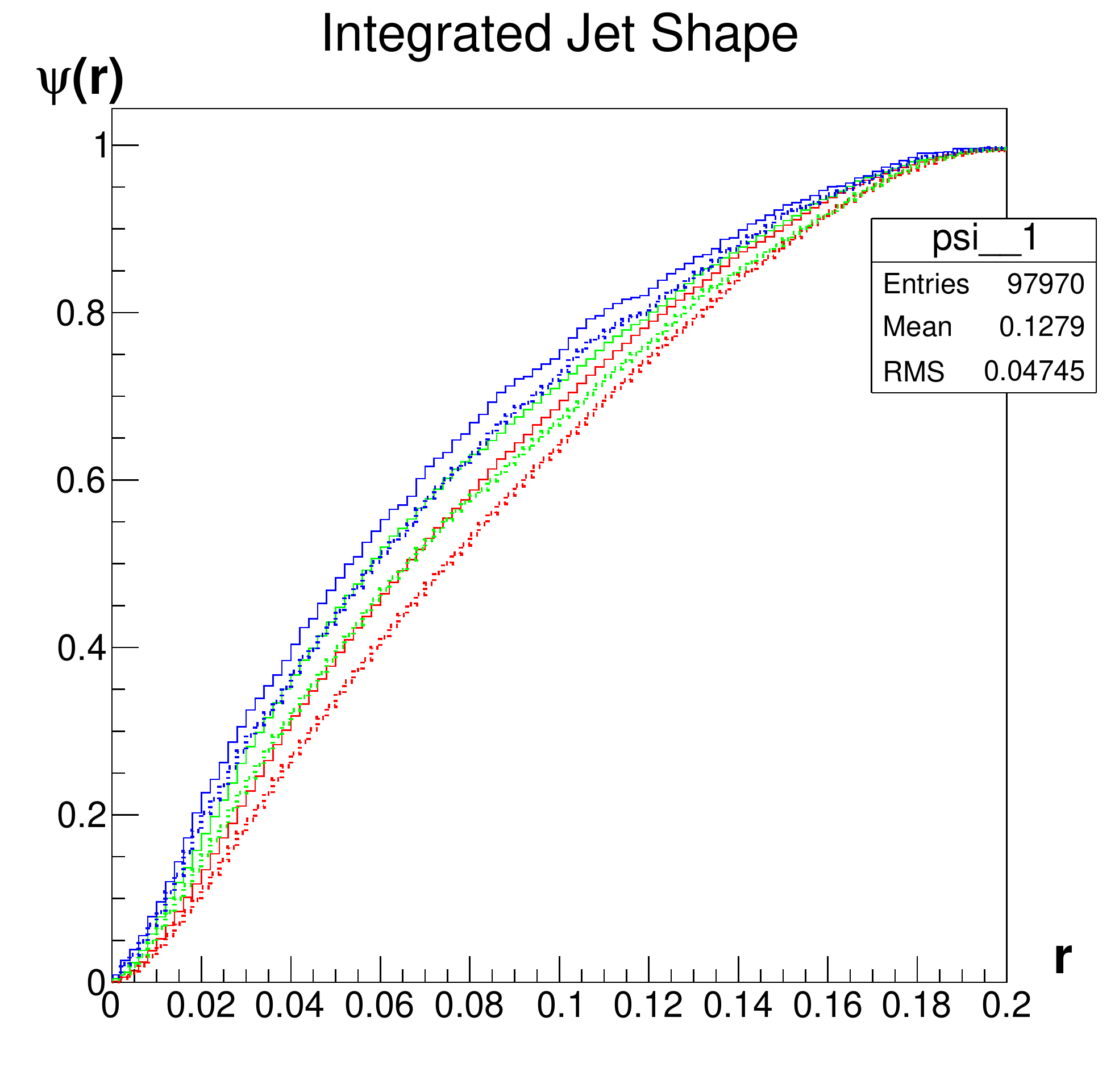}}
\subfloat[$R=0.4$]{\label{fig:collim4}\includegraphics[width=0.35\textwidth]{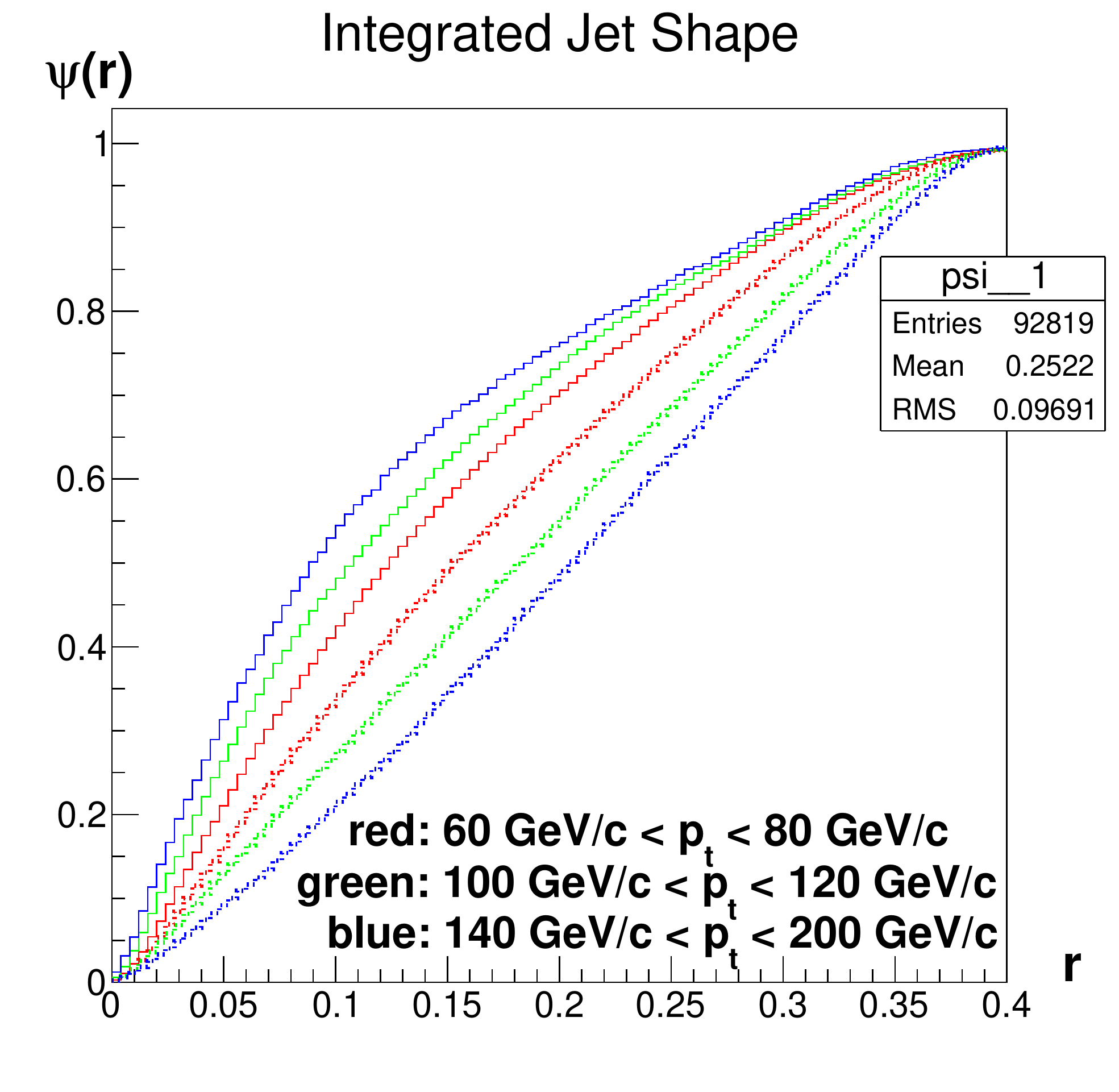}}
\subfloat[$R=0.7$]{\label{fig:collim7}\includegraphics[width=0.35\textwidth]{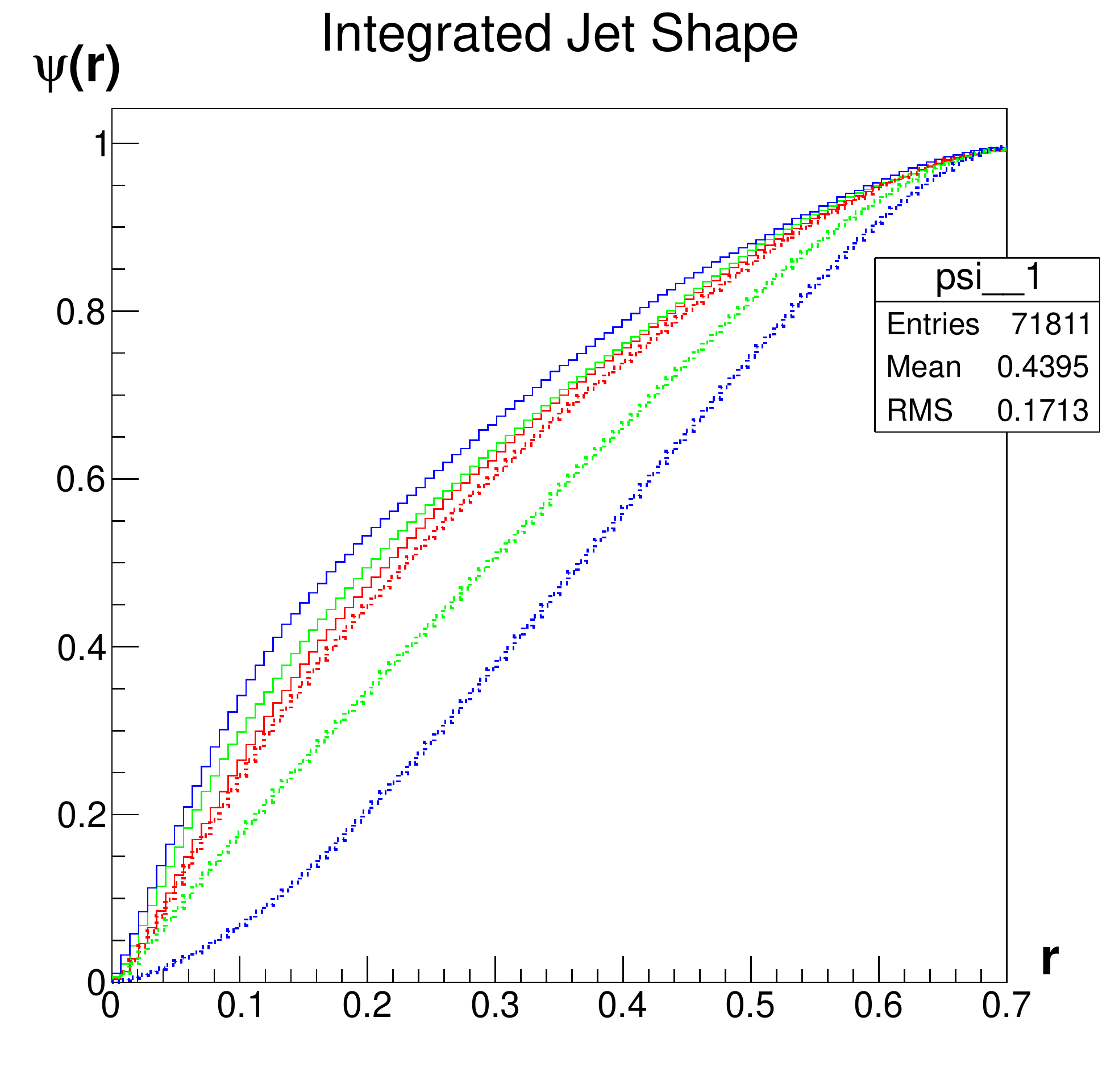}}
}
\caption{Jet collimation for various jet cones. Vacuum (solid), Medium (dotted).}
\label{fig:collimfin}
\end{figure}
\indent With a model roughly replicating the main characteristics of the measured ATLAS data it is of interest to move on to unexplored territory and investigate the intrajet structure resulting from the model. Starting off with the energy distribution within the jet, Fig. \ref{fig:collimtheo} shows the expected consequences of gluon emission in vacuum and medium. Angular ordering ensures that increased gluon emission results in a more collimated jet. Thus, harder jets tend to be more collimated as was seen in Fig. \ref{fig:jetshape} (section \ref{sec:vacuum}). However, in a medium the gluon spreading transfers energy to the edges of the jet cone, or even outside the jet. Hence, jets in a QGP are expected to be less collimated than jets in a vacuum as is shown in the simplest of cases in Fig. \ref{fig:psi}. Note that no adjustment has been made to the angular ordering as implemented by PYTHIA 8. As hard jets are more likely to emit gluons it is expected that these are affected more than less energetic ones. The results are shown in Fig. \ref{fig:collimfin}. \\
\indent The above analysis agrees well with the observations for the intermediate jet cone ($R=0.4$). For the most narrow cone ($R=0.2$) it is clear that the medium has made the jets less collimated. However, it appears that the reduction in $\psi({r})$ is roughly the same for all energy ranges. This is probably related to the fact that most jets are almost completely stripped off of their soft constituents resulting in merely a small spreading of energy inside the cone. The largest jet cone ($R=0.7$) presents more bewildering results. It is clear that the harder jets are feel the medium effects more extensively than softer ones (however not much more than in the case of $R=0.4$). However, it appears as if the softer jets are less affected by the medium than they are for smaller cones. This may be an intrinsic artifact of the model, or possibly related to soft emissions from one cone entering the a neighboring one. \\
\indent Certainly the medium model produces rather large changes compared to the vacuum situation. As an example; considering the hardest jets in the central picture, the central half of the cone carries $\sim 75\%$ of the jet energy in vacuum whilst only $\sim 35\%$ in the  medium. This poses the question whether the magnitude of the transverse momentum boost is of a reasonable size, i.e. what is a realistic value for the parameter $\hat{q}$? \\

\begin{figure}[h]
\begin{center}
\includegraphics[width=4in]{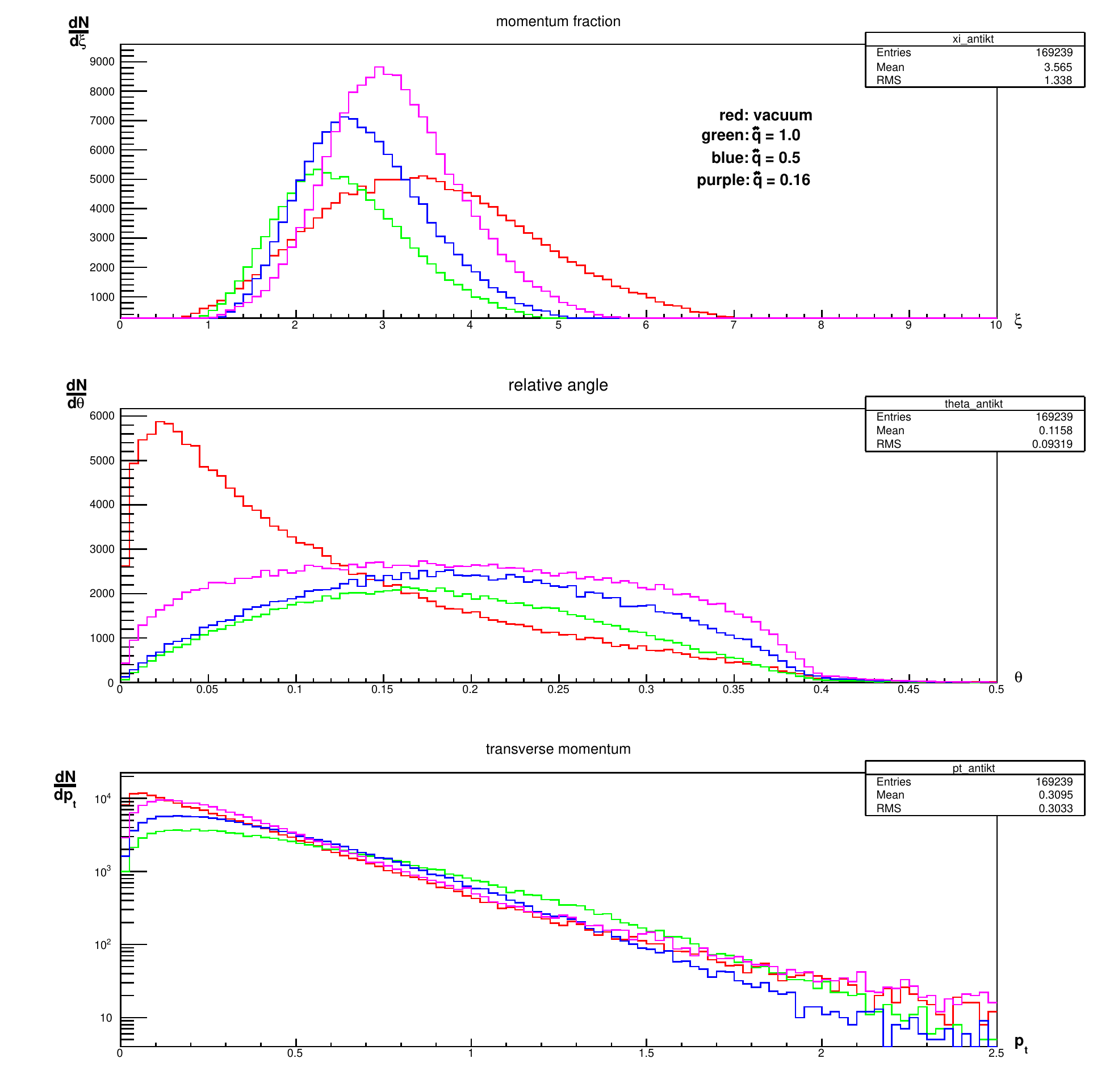}
\caption{Intrajet distributions for various values of $\hat{q}$.}
\label{fig:FFfinal}
\end{center}
\end{figure}

\indent The intrajet distributions in Fig. \ref{fig:FFfinal} demonstrate the differences between various values for $\hat{q}$. The number of particles in the jet is strongly dependent on the transverse momentum assigned to them. For $\hat{q}=0.16$ GeV$^2$/fm (purple) the number of jet constituents have increased compared to the vacuum situation (red). Thus, the modified splitting kernel appears to dominate. However, for $\hat{q}=1.0$ GeV$^2$/fm the number of particles have decreased, thus suggesting that most soft gluons have been pushed outside the jet cone. The modified splitting kernel shows its effect in the reduction of particles with very low $\xi$ (hight $p_t$), whereas the jet collimation demonstrates its presence by reducing the number of particles with very high $\xi$ (soft particles). The precise changes in these features would have to be determined experimentally and would reveal how the two effects balance. \\
\indent The $\theta$-distribution reveals significant differences in the angular distribution of particles within the jet. For $\hat{q}=1.0$ GeV$^2$/fm and $\hat{q}=0.5$ GeV$^2$/fm the distribution forms a rather smeared out hump centered around $\sim 0.16$ and $\sim 0.19$ respectively. For $\hat{q}=0.16$ GeV$^2$/fm the distribution is almost flat with the exception of the extreme edges. This is likely due to the fact that more collisions may occur before the gluon is completely decorrelated from the jet. The jet thus take the shape of only a few hard particles at the center, surrounded by an almost even distribution of soft particles filling up the jet cone to the edge. The $p_t$-distribution demonstrates the same effect. Though most momentum is still carried by the central core particle, this fraction is significantly reduced on behalf of the softer particles that are spread to larger angles.

\section{Conclusion}
This investigation has been aimed at studying how the effects of jet-quenching can be modeled qualitatively. It has been performed by scrutinizing the mathematical theories behind perturbative QCD in the standard model and testing these as reference results. A basic model involving the modification of the Altarelli-Parisi splitting kernels and jet collimation by transversely 'kicking' particles has been attempted and used as a frame for a more serious model accounting for the nuclear geometry. Finally a discussion of the effects of the parameters $f_{med}$ and $\hat{q}$ has been included. \\
\indent The result is a set of distributions for jet energy asymmetry and azimuthal angular separation between the two hardest jets in an event has been obtained. The distributions roughly replicate data collected at the ATLAS experiment in 2010, however with a few discrepancies. Overlooking the defects, the model has then been applied to intrajet distributions and the energy distribution within the jet. The result is a jet which in a medium strongly suffers suppression of its hardest constituents on the behalf of soft gluon emission. These gluons rapidly leave the jet cone leading to a balance between two effects: an increase in gluon production and a successive removal of the same gluons. The balance between these two competing effects is a subject for further theoretical and experimental investigations. Overall, the jet suffers a significant spread of its constituent particles over the entire jet cone. The foremost part of the momentum is centered closely around the jet axis, however the jet energy is increasingly smeared out over the jet cone resulting in a less collimated jet. Each event appears to result in jets of angular separations strongly deviating from the $\pi$ radians commonly observed in a vacuum. Dijet events experience a significant change in asymmetry compared to the vacuum situation. However, events involving more jets appear to be less affected by the medium. Thus we can conclude that the introduction of a quark-gluon plasma not only alters the jet behavior but also distorts the intrinsic structure of the jets. \\
\indent The model presented in the investigation relies on a set of assumptions, primarily that the physics derived for vacuum situations apply in the medium as well. In particular, the fragmentation of the hard process is assumed to be similar to a vacuum. There is no theoretical predictions behind the values $f_{med}$ and $\hat{q}$, and no consideration of longitudinal momentum is included, which would lead to an inelastic energy loss and stopping of the jet fragments in the plasma. The model is based upon a classical treatment of the nuclear geometry and assumes that there is no cross section associated to the area outside the nuclear overlap. A more appropriate approach would be to convolve the Saxon-Woods potentials from the two nuclei. Further the model assumes a stationary, uniform quark-gluon plasma, through which all jets traverse linearly. A more realistic picture would involve an expansion of the plasma with the consequence of reduced temperature and density. This would require a discussion of the thermodynamic properties of the plasma, which has not been a part of this investigation. \\
\indent Despite its discrepancies, the model reproduces some important qualitative features of the ATLAS data. It is useful starting point in the understanding of the underlying reasons for jet properties in a QCD medium. Further, it predicts some interesting results outside of the present experimental scope, which would be equally important to confirm as to disprove. Thus the investigation may be considered successful.


\begin{thebibliography}{99}
\bibitem{Gross} D.J. Gross, F. Wilczek, Phys.Rev.Lett. \textbf{30} 1343-1346 (1973)
\bibitem{Politzer} H.J. Politzer, Phys.Rev.Lett. \textbf{30} 1346-1349 (1973)
\bibitem{Cabibbo} N. Cabibbo, G. Parisi, Phys.Lett.B \textbf{59} 67-69 (1975)
\bibitem{Collins} J. Collins, M.J. Perry, Phys.Rev.Lett. \textbf{34} 1353-1356 (1975)
\bibitem{CERN} http://newstate-matter.web.cern.ch/newstate-matter/Experiments.html
\bibitem{RHIC} I. Arsene, IG Bearden, D. Beavis, C. Besliu, B. Budick, H. B¿ggild, C. Chasman, CH Christensen, P. Christiansen, J. Cibor, et al. Nucl. Phys. A, \textbf{757}(1-2):1Ð27 (2005)
\bibitem{ATLAS} G. Aad \emph{et al.} [ATLAS Collaboration], arXiv:1011.6182v2 [hep-ex] (2010)
\bibitem{CMS} [CMS Collaboration], arXiv:1102.1957v2 [nucl-ex] (2011)
\bibitem{Kniehl} B. A. Kniehl, G. Kramer, B. Potter, Nucl. Phys. B \textbf{582} 514 (2000)
\bibitem{Bethke} S. Bethke, Eur.Phys.J. \textbf{C64} 689 (2009) [arXiv:0908.1135 [hep-ph]]
\bibitem{Hagedorn} R. Hagedorn, Nuovo Cimento Suppl. \textbf{3} 147-186 (1965)
\bibitem{Karsch} F. Karsch, E. Laermann, A. Peikert, Nucl.Phys.B \textbf{605} 579-599 (2001)
\bibitem{Lipatov} V.N. Gribov, L, N. Lipatov. Sov.J.Nucl.Phys. \textbf{15} 438 (1972)
\bibitem{Dok} Yu. L. Dokshitzer. Sov.Phys. JETP, \textbf{46} 641 (1977)
\bibitem{Altarelli} G. Altarelli and G. Parisi. Nucl. Phys. B \textbf{126} 298 (1977)
\bibitem{Salam} Graphics by G. Salam. arXiv:1101.2878v2 [hep-ph] (10 Jan 2011)
\bibitem{book2} R.K. Ellis, W.J Sterling, B.R. Webber, QCD and Collider Physics, ch. 5, Cambride Univertisy Press (1996)
\bibitem{book1} Yu.L. Dokshitzer, V.A. Khoze, A.H. Mueller, S.I. Troyan, Basics of Perturbative QCD, ch. 4, Editions Frontiers (1991)
\bibitem{Lund} B. Andersson, G. Gustafson, G. Ingelman, T. Sjostrand, Phys. Rep. \textbf{97} (1983)
\bibitem{Gavin} Graphics by G. Davies. Imperial College, Lectures on Nucl. and Particle physics (2011)
\bibitem{Sjostrand} T. Sjostrand, S. Mrenna and P. Skands, Comput. Phys. Comm. \textbf{178} (2008) 85 [arXiv:0710.3820]
\bibitem{Cacciari} M. Cacciari, G.P. Salam, G. SoyezPhys. Lett. B \textbf{641} (2006) [hep-ph/0512210]
\bibitem{Dok2} Yu. L. Dokshitzer, V. S. Fadin, V. A. Khoze, Phys. Lett. B \textbf{115} 242 (1982)
\bibitem{Dok3} Yu. L. Dokshitzer, V. S. Fadin, V. A. Khoze, Z. Phys. C \textbf{15} 325 (1982)
\bibitem{Bassetto} A. Bassetto, M. Ciafaloni, G. Marchesini, A. H. Mueller, Nucl. Phys. B \textbf{207} 189 (1982)
\bibitem{PHENIX} K. Adcox \emph{et al.} [PHENIX Collaboration], arXiv:Nucl-ex/0410003 (2005)
\bibitem{STAR} J. Adams \emph{et al.} [STAR Collaboration], arXiv:Nucl-ex/0501009 (2005)
\bibitem{Baier} R. Baier, D. Schiff, B.G. Zakharov, Ann. Rev. Nucl. Part. Sci. \textbf{50}, 37 (2000)
\bibitem{Casalderrey} J. Casalderrey-Solana, J.G. Milhano, U.A. Wiedemann, arXiv:1012.0745v1 [hep-ph] (2010)
\bibitem{Borghini} N. Borghini, U.A. Wiedemann, arXiv:hep-ph/0506218v1 (2005)
\end{thebibliography}
\end{document}